\documentclass[pra, aps, groupedaddress, showpacs, superscriptaddress]{revtex4-2}
\usepackage{amssymb,amsmath,amsthm,color,graphicx,times,graphicx}
\usepackage{hyperref}
\usepackage{subfigure}
\usepackage{times}
\usepackage{bbold}
\usepackage{color}
\usepackage{multirow}
\usepackage{tikz}
\usepackage{amsmath}

\providecommand{\openone}{\leavevmode\hbox{\small1\kern-4.3pt\normalsize1}}

\usepackage{orcidlink}
\usepackage{hyperref}
\hypersetup{
	colorlinks=true,
	linkcolor=blue,
	filecolor=blue,      
	urlcolor=blue,
	citecolor=violet,
	pdfauthor={},
	pdftitle={},
	pdfcreator={Abdallah Slaoui},
}

\usepackage{lipsum}
\usepackage{titlesec}

\theoremstyle{plain}

\theoremstyle{definition}

\begin{document}
\title{Bidirectional quantum teleportation of even and odd coherent states through the multipartite Glauber coherent state: Theory and implementation}
\author{Nada Ikken}\affiliation{LPHE-Modeling and Simulation, Faculty of Sciences, Mohammed V University in Rabat, Rabat, Morocco.}
\author{Abdallah Slaoui \orcidlink{0000-0002-5284-3240}}\email{Corresponding author: abdallah.slaoui@um5s.net.ma}\affiliation{LPHE-Modeling and Simulation, Faculty of Sciences, Mohammed V University in Rabat, Rabat, Morocco.}\affiliation{Centre of Physics and Mathematics, CPM, Faculty of Sciences, Mohammed V University in Rabat, Rabat, Morocco.}
\author{Rachid Ahl Laamara}\affiliation{LPHE-Modeling and Simulation, Faculty of Sciences, Mohammed V University in Rabat, Rabat, Morocco.}\affiliation{Centre of Physics and Mathematics, CPM, Faculty of Sciences, Mohammed V University in Rabat, Rabat, Morocco.}
\author{Lalla Btissam Drissi \orcidlink{0000-0002-1966-9025}}\affiliation{LPHE-Modeling and Simulation, Faculty of Sciences, Mohammed V University in Rabat, Rabat, Morocco.}\affiliation{Centre of Physics and Mathematics, CPM, Faculty of Sciences, Mohammed V University in Rabat, Rabat, Morocco.}\affiliation{College of Physical and Chemical Sciences, Hassan II Academy of Science and Technology, Rabat, Morocco.}
\begin{abstract}
Quantum teleportation has become a fundamental building block of quantum technologies, playing a vital role in the development of quantum communication networks. Here, we present a bidirectional quantum teleportation (BQT) protocol that enables even and odd coherent states to be transmitted and reconstructed over arbitrary distances in two directions. To this end, we employ the multipartite Glauber coherent state, comprising the Greenberger-Horne-Zeilinger, ground and Werner states, as a quantum resource linking distant partners Alice and Bob. The pairwise entanglement existing in symmetric and antisymmetric multipartite coherent states is explored, and by controlling the overlap and number of probes constructing various types of quantum channels, the teleportation efficiency of teleported states in both directions may be maximized. Besides, Alice's and Bob's trigger phases are estimated to explore their roles in our protocol using two kinds of quantum statistical speed referred to as quantum Fisher information (QFI) and Hilbert-Schmidt speed (HSS). Specifically, we show that the lower bound of the statistical estimation error, quantified by QFI and HSS, corresponds to the highest fidelity from Alice to Bob and conversely from Bob to Alice, and that the choice of the pre-shared quantum channel has a critical role in achieving high BQT efficiency. Finally, we show how to implement the suggested scheme on current experimental tools, where Alice can transfer her even coherent state to Bob, and at the same time, Bob can transfer his odd coherent state to Alice.
\end{abstract}
\date{\today}

\maketitle
\section{Introduction}
Quantum teleportation (QT) is an emerging technology for efficiently transferring quantum states. Since it was proposed in 1993, QT has made great progress, both theoretically \cite{Bowen2001,Verstraete2003} and experimentally \cite{JiGang2017,Shengshuai2020}, and has attracted considerable attention \cite{Bennett1992,Bennett1993}. Inspired by Bennett's work, various QT schemes have been suggested to transfer many states and the advantage here is that many of these processes are feasible experimentally, and a lot of them are implemented in reality.
Many demonstrations of these protocols have been made in various physical systems such as electrons \cite{Pfaff2014}, atoms \cite{Krauter2013}, ions \cite{Riebe2004}, photons \cite{Bouwmeester1997}, hybrid systems \cite{Sherson2006,Kirdi2023,Chen2008} and superconducting circuits \cite{Steffen2013}. Exceptional results have been obtained regarding teleportation distance with satellite-based applications on the horizon \cite{Ma2012,Yin2012}. As well, some attempts to extend this to more complex systems have been launched \cite{Wang2015}.\par

In fact, teleportation is understood as a fictitious method, where an object disintegrates in one place and reconstructs itself perfectly in another. Whereas in quantum terms, it is an ability that particles have to share properties and information with each other through quantum systems. If a quantum particle linked to another, regardless of the physical distance from each other, they will be able to exchange information. This concept seems very interesting for everyone, since teleportation is one of the great dreams of humanity and imagined in many works of fiction. It seems that in quantum terms, it has been achieved with Qubits, however, the areas that can benefit from QT are logically the secure communications that are already used. A major feature to consider here would be that the phenomenon producing QT is quantum entanglement \cite{Einstein1935,Bell1966,Hill1997}, known as quantum correlations \cite{Maziero2009,SlaouiS2018,Slaoui2018}. Which in turn, is defined as a particular interrelation between objects, i.e., the measurement of one object immediately influences the other even though the two remain entirely isolated and separate from each other. Therefore, the concept of entanglement is one of the most important properties of the deep knowledge of quantum mechanics, due to its important significance not only for QT, but also for quantum computing \cite{Steane1998,Ladd2010}, quantum dense coding \cite{Mattle1996} and quantum key distribution \cite{Ekert1991}.\par

Broadly speaking, Alice (a sender) and Bob (a receiver) must share a resource state, which must be an entangled quantum state, for the transfer of an unknown quantum state to be performed. Throughout the teleportation procedure, the resource state should remain entangled. For Alice to exchange measurement results with Bob, both Alice and Bob must have a communication channel. Superluminal communication is not allowed in this process, but it requires classical communication and a shared entanglement resource \cite{Bennett1992}. Furthermore, only when the resource state is maximally entangled and the system is devoid of decoherence and dissipation can Bob employ unitary transformations to recover an exact copy of the input state \cite{Kirdi2022,KirdiSlaoui2023}. In multiparty quantum teleportation, several studies have been done. Hitherto, several QT schemes have been successively proposed using various entangled quantum states. To the best of our knowledge, Karlsson et al.\cite{Karlsson1998} suggested the first QT between three participants utilizing the GHZ state. Dong and his colleagues \cite{Dong2011} carried out a controlled communication between the three parties utilizing imperfect Bell state measurement and GHZ state. Furthermore, Hassanpour and his collaborators use GHZ-like states to realize a controlled quantum secure for direct communication protocol \cite{Hassanpour2015}. It is essential to consider scenarios with low entanglement and reduced efficiency because, under realistic conditions, the effect of the environment reduces the degree of entanglement between the system components. From this, the concept of probabilistic QT with high probability of success was born \cite{Agrawal2002}. Further, the standard QT protocol has been extended to bidirectional (two-way) quantum teleportation (BQT), where Alice and Bob transmit their states simultaneously to each other using preshared entanglement, and it is an integrated feature of QT networks \cite{Vaidman1994,Kiktenko2016}. Along these lines, Fu et al.,\cite{Fu2014} reported a BQT scheme using a four-qubit cluster state as a quantum channel where Alice and Bob can exchange single-qubit states via the Hadamard operation, specified unitary operations, and Bell basis measurement in their system concurrently. Then, Zha et al.,\cite{Zha2013} proposed a bidirectional controlled QT protocol where Alice and Bob can simultaneously send unknown quantum states to one another under the supervision of all controllers.\par

Recent progress has shown that coherent states are playing a significant role in many applications of QT and communication, which are easily produced and manipulated using optical components \cite{Schrodinger1926,Glauber1963,Sudarshan1963}. A variety of schemes have been proposed for generating such superposition states as well as their entangled counterparts using nonlinear interactions, photon subtraction from squeezed vacuum states, mixing of squeeze vacuum with a coherent light \cite{Yurke1986}. There were also suggestions for creating freely moving multiparty resources such as GHZ, W and cluster entangled coherent states \cite{An2009}. Since then, the field has continued to grow, as superposed coherent states can offer several advantages in practice, including its robustness against noisy channels and its status as the fastest method of information transmission \cite{Jeong2001}. In this direction, Van-Enk et al.,\cite{Enk2001} proposed a standard quantum teleportation scheme of a single qubit encoded in coherent state superpositions $\vert I\rangle = \epsilon_{+} \vert\alpha\rangle +\epsilon_{-} \vert -\alpha\rangle$. The scheme's success probability was demonstrated to be $50\%$. Nevertheless, their suggested plan has a few flaws. For instance, mixing modes that are located at various locations using a beam splitter, such as one mode that is with Alice and the other mode that is with Bob. Besides, Pyrkov and Byrnes \cite{Pyrkov2014} suggested a new scheme for teleporting a coherent spin state between two-component Bose-Einstein condensates (atomic ensembles) and they found that the error of the proposed protocol scales favorably with the number particles in the coherent spin states, which reaches fidelities and are close to $100\%$ in comparison with the all-classical strategy. This protocol has been generalized by the same authors \cite{Pyrkov22014} to teleport a spin coherent state to another state of the same type with an arbitrary position on the Bloch sphere. Motivated by these works, a new BQT scheme is proposed here to transmit arbitrary even and odd coherent states in both ways between the two users (i.e., bidirectional teleported states). Using the quantum channel composed of multipartite Glauber coherent states, user Alice can transmit an even coherent state to user Bob, and simultaneously, user Bob can transmit an odd coherent state to user Alice \cite{Furusawa2011,Dodonov1974,Gou2017}.\par
Mathematically, the efficiency of the QT process is measured by the fidelity susceptibility between the input and output states \cite{Uhlmann1976,You2007}. This concept is closely related to the quantum Fisher information (QFI) such that the expression of the fidelity susceptibility is proportional to the expression of QFI for any teleported state \cite{Hubner1992,Braunstein1994,Zanardi2008}. Therefore, it is feasible to estimate the teleported and gained parameters using the QFI which plays a primary role in quantum estimation through the quantum Cramer-Rao bound \cite{Giovannetti2006,Helstrom1969,Bakmou2019}. Indeed, quantum metrology aims at improving the accuracy of measurements by reducing their fundamental statistical uncertainty given by quantum fluctuations. This is achieved by preparing the physical system in a state that has particular quantum correlations. A fundamental tool for evaluating and comparing the effectiveness of various metrological strategies for quantum parameter estimates is provided by the QFI, where the higher the QFI, the higher the precision \cite{Paris2009,Shaji2007,Abouelkhir2023}. Based on the knowledge that the QFI can be derived as a quantum statistical speed from the Hellinger distance \cite{Jeffreys1946}, the Hilbert-Schmidt speed has in turn been used to improve the precision of the estimated parameters that detects the lower limit of the statistical estimation error like QFI \cite{Gessner2018}. Moreover, it has the advantage of avoiding the diagonalization of the evolved density matrix and would be useful for high-dimensional quantum systems.\par

Here, we implement BQT using two independent quantum random trigger qubits, one of which corresponds to the even coherent state and the other to the odd coherent state. The global state of the system and an entangled state between the even and odd states (i.e. an entangled resource connecting Alice and Bob) are then combined to create a circuit that was employed in the Qiskit (quantum computer), which ultimately gives us a measurement of the probabilities. Then, in order to overcome the traditional fidelity barrier, we derive the amount of shared state entanglement quantified by Wootters concurrence, the teleportation fidelity, the success probability and estimate the weight parameters of the teleported states. We show how the trigger states and initial states of the transported qubits affect the fidelity and QFI of bidirectional teleportation, where both the maximum QFI and fidelity values are the same even when predicted at distinct polarization angles. The HSS was determined using QFI, and it was shown how well this statistical speed quantifier can be applied to the BQT protocol. Further, by optimizing the entanglement resources, we can achieve high-fidelity teleportation even in the presence of noise and decoherence. The even and odd coherent states form the basis of our protocol and enable the transmission of qubit states bidirectionally using a single entangled state and two separate trigger qubits. The entanglement resources are carefully chosen to ensure that the teleported states are faithfully reproduced at the receiving end. Finally, we conduct simulations using Qiskit to test the proposed BQT protocol in a controlled environment and their capacity to transfer information bidirectionally. By testing the protocol in a simulated environment, we can identify potential issues and optimize the protocol accordingly. Once the simulations are successful, we can move on to implementing the protocol on a real quantum computer.
 
\section{Multipartite Coherent states in BQT protocol}\label{Sec2}
\subsection{Multipartite entanglement in Even and Odd coherent states}
Any quantum technology attempts to exploit the non-classical properties of light and matter to achieve supremacy and potentially exponential efficiencies over classical deterministic approaches. There are many types of states in this regard, the most typical of which is the coherent state. Originally, the last state were proposed by Schrödinger for the quantum harmonic oscillator as specific quantum states whose properties are more similar to those of their classical counterpart \cite{Schrodinger1926}. Such states are regarded as quasi-classical states, minimizing the Heisenberg uncertainty for the position and momentum operators, and maintaining maximum localizability throughout the temporal evolution of the system. Thence, in the 1960s, Glauber \cite{Glauber1963,Glauber21963}, Sudarshan \cite{Sudarshan1963} and Klauder \cite{Klauder1963,Klauder21963} launched this fruitful field by showing the applicability and relevance of these states in quantum optics. Theoretically, Glauber proved that these states can be defined as eigenvectors of the bosonic annihilation operator and can be obtained by applying a displacement operator ${\cal D}\left(\eta\right)=\exp \left(\eta a^{\dagger}-\eta^{*} a\right)$ to the vacuum state of the harmonic oscillator as
\begin{equation}
|\eta\rangle\equiv D(\eta)|0\rangle=\exp\left[-\frac{|\eta|^{2}}{2}\right]\sum_{n} \frac{\eta^{n}}{\sqrt{n !}}|n\rangle,
\end{equation}
where $\eta$ is the complex dimensionless amplitude of the coherent state and $|n\rangle$ stands for the Fock state. When the concept is applied to a quantized monochromatic electromagnetic wave, $n$ represents the number of elementary quanta of energy, known as photons. On the other hand, the entire Hilbert space is a tensor product for many systems of $n$ non-interacting particles. This enables the superposition of tensorial products with the form $\left|\eta_{1}\right\rangle \otimes\left|\eta_{2}\right\rangle\otimes\ldots\otimes\left|\eta_{n}\right\rangle \equiv\left|\eta_{1}\eta_{2}\ldots \eta_{n}\right\rangle$ to be used to represent any multipartite coherent state. To make our goal clearer, we will focus mainly on the superposition of multipartite coherent states between $\left|\eta\right\rangle$ and $\left|-\eta\right\rangle$ as
\begin{equation}\label{eq2}  
	|\psi\rangle\equiv\left|\eta,m,n\right\rangle\simeq{\cal N}\left(|\eta, \ldots, \eta\rangle+e^{i m \pi}|-\eta, \ldots,-\eta\rangle\right),
\end{equation}
where the normalization factor ${\cal N}$ reads
\begin{equation}
{\cal N}=\left[2+2p^{n}\cos m\pi \right]^{-1/2},
\end{equation}
with $m\in  {\mathbb Z}$, $p=\left\langle\eta\mid-\eta\right\rangle =\exp\left(-2\mid\eta\mid^{2} \right)$ is the overlap between the states, $m=1$ for antisymmetric states and $m=0$ for symmetric states. As it turns out, we will mainly investigate the two asymptotic bounds of multipartite coherent states, with one case being $p\rightarrow0$ and the second case being $p\rightarrow1$. In the first limiting situation $p\rightarrow0$, the two coherent states $\left|\eta\right\rangle$ and $\left|-\eta\right\rangle$ are orthogonal and constitute an orthogonal basis $\left\lbrace\left|0\right\rangle\equiv\left|\eta\right\rangle,\left|1\right\rangle\equiv\left|-\eta\right\rangle \right\rbrace$. Therefore, ${\cal N}=1/2$ and the state $\left|\eta,m,n\right\rangle$ acts as a multipartite state of Greenberger-Horne-Zeilinger (GHZ) type such as
\begin{align}
\left|\eta,m,n\right\rangle\simeq\left|{GHZ} \right\rangle_{n}=&\frac{1}{\sqrt{2}}\left[\left|\eta\right\rangle \otimes\left|\eta\right\rangle\otimes\ldots\otimes\left|\eta\right\rangle +e^{im\pi}\left|-\eta\right\rangle \otimes\left|-\eta\right\rangle\otimes\ldots\otimes\left|-\eta\right\rangle \right].  
\end{align}
Furthermore, the situations where $m=0$ and $m=1$ must be distinguished individually in the second limiting case where $p\rightarrow1$ (or $\eta\rightarrow0$). For symmetric states ($m$ even), the multipartite coherent state (\ref{eq2}) reduces to the ground state as
\begin{align}
\left|\eta,0,n\right\rangle\simeq\left|\eta\right\rangle \otimes\left|\eta\right\rangle\otimes\ldots\otimes\left|\eta\right\rangle,
\end{align}
and for the second situation where $m$ odd (i.e., for the antisymmetric states), the state $\left|\eta,1,n\right\rangle$ can be observed as a multipartite state of Werner-type, i.e.,
\begin{align}
\left|\eta,1,n\right\rangle\simeq\left|W \right\rangle_{n}=\frac{1}{\sqrt{n}}&\left[\left|-\eta\right\rangle \otimes\left|\eta\right\rangle\otimes\ldots\otimes\left|\eta\right\rangle+ \left|\eta\right\rangle \otimes\left|-\eta\right\rangle\otimes\ldots\otimes\left|\eta\right\rangle+\ldots +\left|\eta\right\rangle \otimes\left|\eta\right\rangle\otimes\ldots\otimes\left|-\eta\right\rangle\right].
\end{align}
In other words, the symmetric states interject between the GHZ-like states ($p\rightarrow0$) and the separable state ($p\rightarrow1$). Conversely, we observe that the antisymmetric states are interposed between the Werner-type states ($p\rightarrow1$) and the GHZ-type states ($p\rightarrow0$). Here, we separated our system into two parts to be able to use it in quantum teleportation. The one generates mixed coherent states, while the other generates exclusively pure coherent states. As a first bipartite splitting of multipartite coherent states, we formulate Eq.(\ref{eq2}) as follows 
\begin{equation}\label{eq3} 
|\psi\rangle=N\left(|\eta\rangle_{r} \otimes|\eta\rangle_{n-r}+e^{i m \pi}|-\eta\rangle_{r} \otimes|-\eta\rangle_{n-r}\right)
\end{equation}
where $|\pm \eta\rangle_{k}=|\pm \eta\rangle_{1} \otimes|\pm \eta\rangle_{2} \otimes \cdots \otimes|\pm \eta\rangle_{k}$, and $k=r,n-r$. In fact, the total system is divided into two subsystems, one containing any $r$ particles ($1\leq r\leq n-1$) and the other containing all the remaining $n-r$ particles. Two logical qubits can be used to represent the multi-particle state (2) and the implementation of this is given by
\begin{equation}\label{eq4} 
	|0\rangle_{r}=\frac{|\eta\rangle_{r}+|-\eta\rangle_{r}}{\sqrt{2\left(1+p^{r}\right)}},\hspace{1cm}|1\rangle_{r}=\frac{|\eta\rangle_{r}-|-\eta\rangle_{r}}{\sqrt{2\left(1-p^{r}\right)}},
\end{equation}
for the first subsystem in the orthogonal basis $\left\{|0\rangle_{r},|1\rangle_{r}\right\}$, and likewise
\begin{equation}\label{eq5}  
|0\rangle_{n-r}=\frac{|\eta\rangle_{n-r}+|-\eta\rangle_{n-r}}{\sqrt{2\left(1+p^{n-r}\right)}}, \hspace{1cm}|1\rangle_{n-r}=\frac{|\eta\rangle_{n-r}-|-\eta\rangle_{n-r}}{\sqrt{2\left(1-p^{n-r}\right)}},
\end{equation}
for the second subsystem in the basis $\left\{|0\rangle_{n-r},|1\rangle_{n-r}\right\}$. Inserting the equations (\ref{eq4}) and (\ref{eq5}) into (\ref{eq3}), the pure bipartite quantum state (\ref{eq3}) in the computational basis $\left\{|0\rangle_{r} \otimes|0\rangle_{n-r},|0\rangle_{r} \otimes|1\rangle_{n-r},|1\rangle_{r} \otimes|0\rangle_{n-r},|1\rangle_{r} \otimes|1\rangle_{n-r}\right\}$ can be easily rewritten as
\begin{equation}
|\psi\rangle=\sum_{\alpha=0,1} \sum_{\beta=0,1} G_{\alpha, \beta}|\alpha\rangle_{r} \otimes|\beta\rangle_{n-r},
\end{equation}
with the factors $G_{\alpha, \beta}$ are given by
\begin{align}
&G_{0,0}={\cal N}\left(1+e^{i m \pi}\right) a_{r} a_{n-r}, \hspace{1cm} G_{0,1}={\cal N}\left(1-e^{i m \pi}\right) a_{r} b_{n-r}\notag\\&
G_{1,0}={\cal N}\left(1-e^{i m \pi}\right) a_{n-r} b_{r}, \hspace{1cm} G_{1,1}={\cal N}\left(1+e^{i m \pi}\right) b_{r} b_{n-r},
\end{align}
where the quantities $a_{k}$ and $b_{k}$ in terms of the overlap $p$ are 
\begin{align*}
a_{k}=\sqrt{\frac{1+p^{k}}{2}}, \hspace{0.5cm}{\rm and}\hspace{0.5cm} b_{k}=\sqrt{\frac{1-p^{k}}{2}},\hspace{0.5cm} k=r,n-r.
\end{align*}
Further, the second bi-partition $\rho_{rk}$ can be produced from the complete $n$-particles system by tracing all particles except those specified by the indices $r$ and $k$. More exactly, there are $n(n-1)/2$ dissimilar density matrices with the notation $rk$. In essence, all the reduced density matrices $\rho_{rk}$ are the same. The next step is to establish $\rho_{12}$ and generalize it. Such a reduced density matrix is obtained as
\begin{equation}\label{eq9}
\begin{aligned}
\rho_{12}&= \operatorname{Tr}_{3,4, \ldots, n}\left(|\psi\rangle\langle\psi|\right) \\
& =\mathcal{N}^{2}\left[|\eta, \eta\rangle\left\langle\eta, \eta\left|+e^{-i m \pi} p^{n-2}\right| \eta, \eta\right\rangle\langle-\eta,-\eta|+e^{i m \pi} p^{n-2}|-\eta,-\eta\rangle\langle\eta, \eta|+|-\eta,-\eta\rangle\langle-\eta,-\eta|\right].
\end{aligned}
\end{equation}
It's important to notice that the states $|\pm \eta\rangle$ do not serve as an orthogonal basis for the Fock Hilbert space $\mathcal{H}$. Moreover, the coherent states are usually non-orthogonal; in fact, their overlap is $|\langle\eta \mid \delta\rangle|^{2}=\exp\left(-|\eta-\delta|^{2}\right)$ for the two coherent states $|\eta\rangle$ and $|\delta\rangle$. Nevertheless, an orthonormal basis can be obtained by taking the even and odd coherent states \cite{Ansari1994,Dodonov1974} which are defined by
\begin{align}
	&\left|Even, \eta_{e}\right\rangle\equiv\left|\eta_{e}\right\rangle=N_{+}\left(\left|\eta\right\rangle+\left|-\eta\right\rangle\right),\hspace{2cm} \left|Odd,\eta_{o}\right\rangle\equiv\left|\eta_{o}\right\rangle=N_{-}\left(\left|\eta\right\rangle-\left|-\eta\right\rangle\right),\label{Eq10}
\end{align}
where the normalization factors $N_{\pm}$ are given by
\begin{align*}
	N_{\pm}=\left[2\left(1\pm p\right)\right]^{-\frac{1}{2}}.
\end{align*}
The logical qubits involved in the two-dimensional Hilbert space are encrypted in terms of even and odd coherent states as $\left|{\bf0}\right\rangle_{L}\longmapsto\left|\eta_{e}\right\rangle$ and $\left|{\bf1}\right\rangle_{L}\longmapsto\left|\eta_{o}\right\rangle$. Then one can show that 
\begin{align}
&\left|\eta\right\rangle\equiv {\bf a}\left|{\bf0}\right\rangle_{L}+{\bf b}\left|{\bf1}\right\rangle_{L},\hspace{2cm}\left|-\eta\right\rangle\equiv {\bf a}\left|{\bf0}\right\rangle_{L}-{\bf b}\left|{\bf1}\right\rangle_{L},
\end{align}
with
\begin{equation}
{\bf a}=\sqrt{\frac{1+p}{2}},\hspace{1cm}{\rm and}\hspace{1cm}{\bf b}=\sqrt{\frac{1-p}{2}}.
\end{equation}
As a result, it is straightforward to verify that the superposition of multipartite coherent states (\ref{eq9}) can be written in the orthonormal basis (\ref{Eq10}) formed by the even and odd coherent states $\left\lbrace \left|\eta_{e}\eta_{e}\right\rangle,\left|\eta_{e}\eta_{o}\right\rangle,\left|\eta_{o}\eta_{e}\right\rangle,\left|\eta_{o}\eta_{o}\right\rangle\right\rbrace $ as
\begin{align}\label{rhoij}
\rho_{12}&=2\mathcal{N}^{2}\left\lbrace \left({\bf a}^{4}+{\bf b}^{4} \right)\left(1+p^{n-2}\cos m\pi \right)\left[\left|\eta_{e}\eta_{e}\right\rangle \left\langle\eta_{e}\eta_{e}\right|+\left|\eta_{o}\eta_{o}\right\rangle \left\langle\eta_{o}\eta_{o}\right|\right] +{\bf a}^{2}{\bf b}^{2}\left(1-p^{n-2}\cos m\pi \right)\right.\notag\\& \left.\left[\left|\eta_{e}\eta_{o}\right\rangle \left\langle\eta_{e}\eta_{o}\right|+\left|\eta_{e}\eta_{o}\right\rangle \left\langle\eta_{o}\eta_{e}\right|+\left|\eta_{o}\eta_{e}\right\rangle \left\langle\eta_{e}\eta_{o}\right|+\left|\eta_{o}\eta_{e}\right\rangle \left\langle\eta_{o}\eta_{e}\right|\right]+{\bf a}^{2}{\bf b}^{2}\left(1+p^{n-2}\cos m\pi \right)\left[\left|\eta_{e}\eta_{e}\right\rangle \left\langle\eta_{o}\eta_{o}\right|+\left|\eta_{o}\eta_{o}\right\rangle \left\langle\eta_{e}\eta_{e}\right|\right]\right\rbrace.
\end{align}
Actually, the above states (\ref{rhoij}) holds an intrinsic amount of entanglement between its parts. One of the easiest ways to measure quantum entanglement is through concurrence \cite{Wootters1998}, which is defined as
\begin{equation}
\mathcal{C}(\rho)=\max \left\{0, \sqrt{\lambda_{1}}-\sqrt{\lambda_{2}}-\sqrt{\lambda_{3}}-\sqrt{\lambda_{4}}\right\},
\end{equation}
where $\lambda_{i}$ are the eigenvalues of the Hermition matrix $\rho\tilde{\rho}$ in decreasing order, with $\tilde{\rho}=\left(\sigma_{y}\otimes\sigma_{y}\right)\rho^{*}\left(\sigma_{y}\otimes\sigma_{y}\right)$ is the spin-flip density matrix. Using Wootters formula for concurrence, the explicit expression of the concurrence entanglement in the state (\ref{rhoij}) can be expressed as
\begin{equation}
\mathcal{C}\left(\rho_{12}\right)=\max\left\lbrace0,\mathcal{C}_{+},\mathcal{C}_{-}\right\rbrace,
\end{equation}
with
\begin{align}
\mathcal{C}_{\pm}=\frac{1-p^{2}}{2\left(1+p^{n}\cos m\pi \right)}&\left[\mid\left(1\pm p^{n-2}\cos m\pi \right)\mid-\left(1\mp p^{n-2}\cos m\pi \right) \right]. 
\end{align}
\begin{figure}[hbtp]
	{{\begin{minipage}[b]{.45\linewidth}
				\centering
				\includegraphics[scale=0.4]{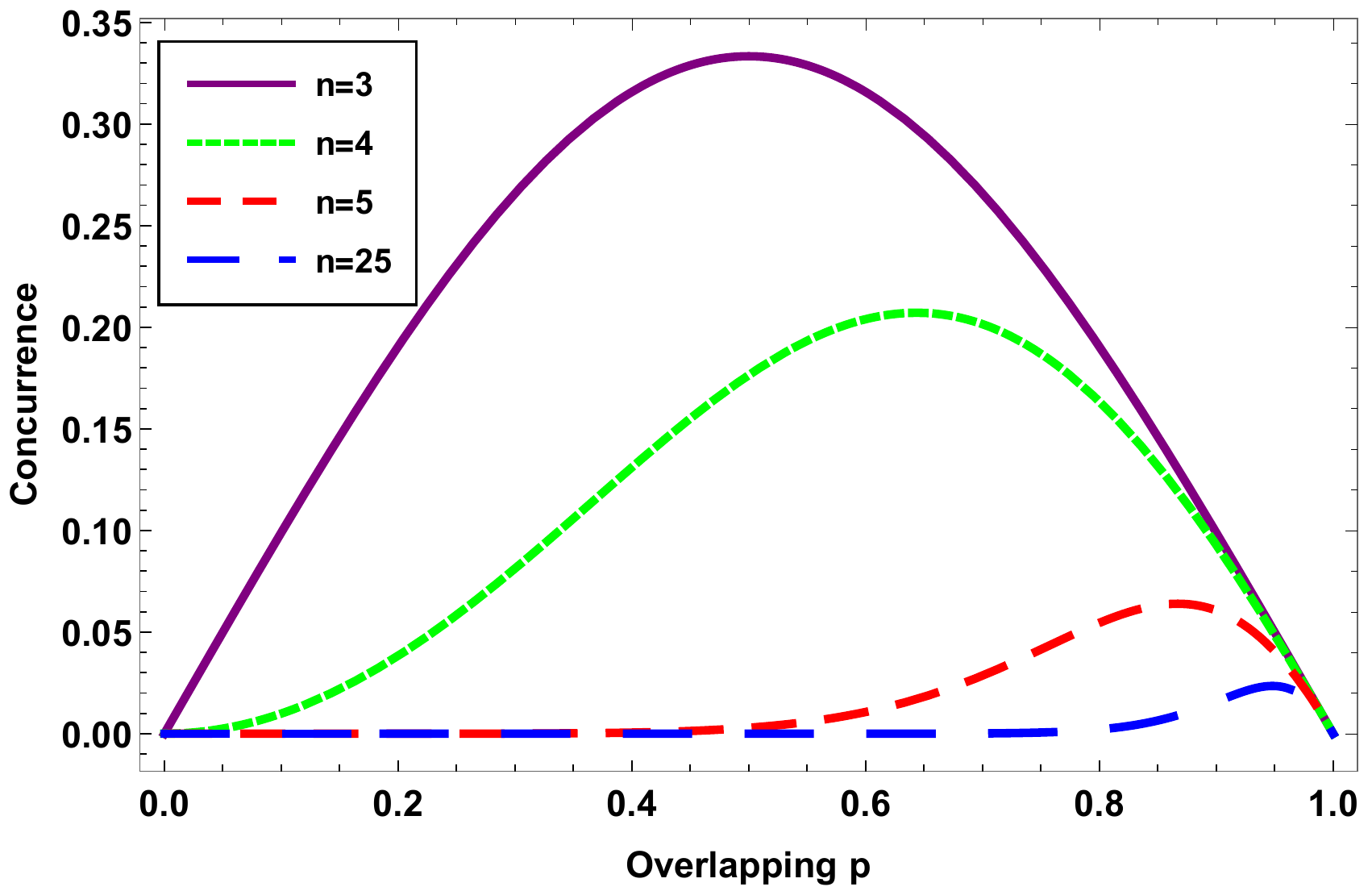} \vfill $\left(a\right)$ For symmetric states
			\end{minipage}
			\begin{minipage}[b]{.45\linewidth}
				\centering
				\includegraphics[scale=0.4]{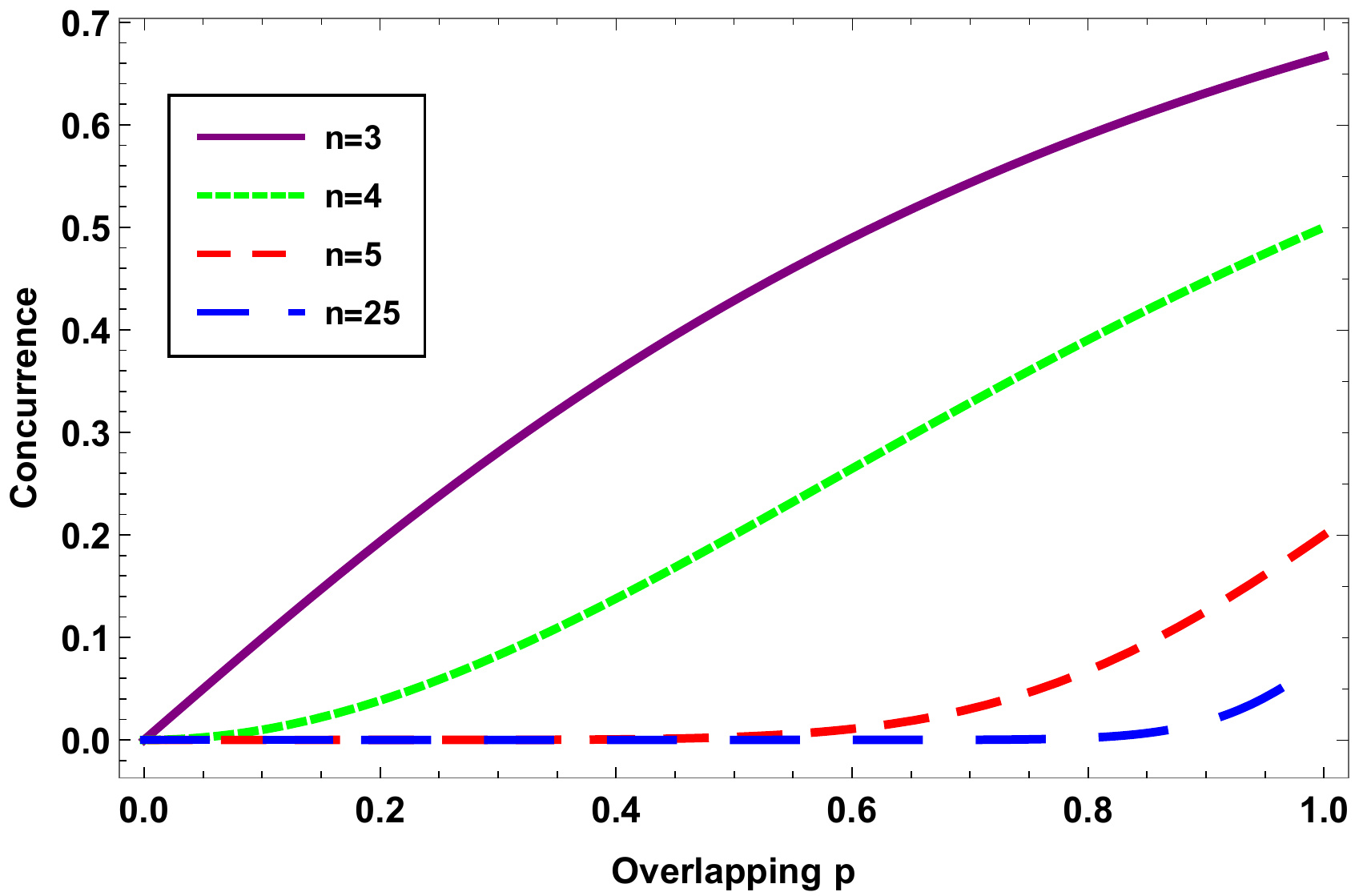} \vfill  $\left(b\right)$ For antisymmetric states
	\end{minipage}}}
	\caption{Behavior of concurrence entanglement as a function of the coherent state overlapping $p$ for different values of the number of probes $n$; $(a)$ for symmetric multipartite coherent states ($m$ even), $(b)$ for antisymmetric multipartite coherent states ($m$ odd).}\label{Fig1}
\end{figure}
In order to analyze the influence of both the coherent state overlapping and the number of probes on the entangled resource linking sender and receiver in our BQT protocol, we plot in Fig.(\ref{Fig1}) the behavior of the concurrence entanglement of multipartite coherent states as a function of the overlap $p$ for some values of $n$. As sketched in Fig.\ref{Fig1}($a$) for symmetric coherent states ($m$ even), after the initial increase, the collective entanglement between the two parts of multipartite coherent states decreases to disappear in the limiting case $p\rightarrow1$, in which the resource states (\ref{eq2}) become separable states. Moreover, we observe that the increase in the number of particles leads to a decrease in the degree of entanglement. In Fig.\ref{Fig1}($b$), we depict the degree of entanglement for antisymmetric multipartite coherent states ($m$ odd). The concurrence entanglement increases with increasing values of the overlap $p$ and reaches its maximum value in the limiting case $p\rightarrow1$, where the state (\ref{eq2}) reduces to a Werner-type multipartite state. In the limit $p\rightarrow0$ and also for the large number of particles, Fig.\ref{Fig1}($b$) reveals that there is no entanglement between the two parts forming the two sub-systems of coherent states. These results indicate that the probe numbers $n$ and the overlap $p$ of the multipartite coherent states are factors that could be controlled to achieve pre-shared entanglement (i.e., quantum resource connecting Alice and Bob) in our BQT protocol.

\subsection{Description of the proposed BQT protocol using Even and Odd coherent states}
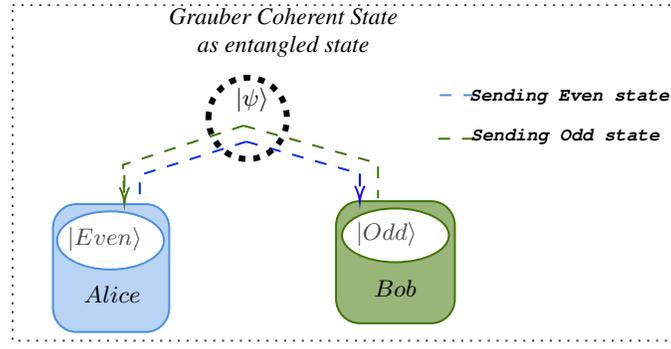
\begin{figure}
	\tikzset{every picture/.style={line width=0.75pt}} 
	\begin{tikzpicture}[x=0.53pt,y=0.55pt,yscale=-1,xscale=1]
		\draw  [color={rgb, 255:red, 74; green, 144; blue, 226 }  ,draw opacity=1 ][fill={rgb, 255:red, 74; green, 144; blue, 226 }  ,fill opacity=0.39 ] (148.8,210.6) .. controls (148.8,201.43) and (156.23,194) .. (165.4,194) -- (215.2,194) .. controls (224.37,194) and (231.8,201.43) .. (231.8,210.6) -- (231.8,265.4) .. controls (231.8,274.57) and (224.37,282) .. (215.2,282) -- (165.4,282) .. controls (156.23,282) and (148.8,274.57) .. (148.8,265.4) -- cycle ;
		\draw  [color={rgb, 255:red, 65; green, 117; blue, 5 }  ,draw opacity=1 ][fill={rgb, 255:red, 65; green, 117; blue, 5 }  ,fill opacity=0.53 ] (350.84,208.8) .. controls (350.84,199.52) and (358.36,192) .. (367.64,192) -- (418.84,192) .. controls (428.12,192) and (435.64,199.52) .. (435.64,208.8) -- (435.64,259.2) .. controls (435.64,268.48) and (428.12,276) .. (418.84,276) -- (367.64,276) .. controls (358.36,276) and (350.84,268.48) .. (350.84,259.2) -- cycle ;
		\draw [color={rgb, 255:red, 9; green, 46; blue, 232 }  ,draw opacity=1 ] [dash pattern={on 4.5pt off 4.5pt}]  (286.8,151) -- (210.8,174) -- (210.8,193) ;
		\draw [color={rgb, 255:red, 9; green, 46; blue, 232 }  ,draw opacity=1 ] [dash pattern={on 4.5pt off 4.5pt}]  (286.8,151) -- (367.8,173) -- (367.8,192) ;
		\draw [color={rgb, 255:red, 26; green, 9; blue, 224 }  ,draw opacity=1 ] [dash pattern={on 4.5pt off 4.5pt}]  (367.8,173) -- (367.8,182) -- (367.8,190) ;
		\draw [shift={(367.8,192)}, rotate = 270] [color={rgb, 255:red, 26; green, 9; blue, 224 }  ,draw opacity=1 ][line width=0.75]    (10.93,-3.29) .. controls (6.95,-1.4) and (3.31,-0.3) .. (0,0) .. controls (3.31,0.3) and (6.95,1.4) .. (10.93,3.29)   ;
		\draw [color={rgb, 255:red, 65; green, 117; blue, 5 }  ,draw opacity=1 ] [dash pattern={on 4.5pt off 4.5pt}]  (284.8,140) -- (380.8,166) -- (380.8,180) -- (380.8,189.99) ;
		\draw [color={rgb, 255:red, 65; green, 117; blue, 5 }  ,draw opacity=1 ] [dash pattern={on 4.5pt off 4.5pt}]  (284.8,140) -- (198.8,167) -- (199.8,193.01) ;
		\draw [color={rgb, 255:red, 65; green, 117; blue, 5 }  ,draw opacity=1 ] [dash pattern={on 4.5pt off 4.5pt}]  (199.8,174.01) -- (199.8,183.01) -- (199.8,191.01) ;
		\draw [shift={(199.8,193.01)}, rotate = 270] [color={rgb, 255:red, 65; green, 117; blue, 5 }  ,draw opacity=1 ][line width=0.75]    (10.93,-3.29) .. controls (6.95,-1.4) and (3.31,-0.3) .. (0,0) .. controls (3.31,0.3) and (6.95,1.4) .. (10.93,3.29)   ;
		\draw  [dash pattern={on 2.53pt off 3.02pt}][line width=2.25]  (259.8,135) .. controls (259.8,119.54) and (272.34,107) .. (287.8,107) .. controls (303.26,107) and (315.8,119.54) .. (315.8,135) .. controls (315.8,150.46) and (303.26,163) .. (287.8,163) .. controls (272.34,163) and (259.8,150.46) .. (259.8,135) -- cycle ;
		\draw  [color={rgb, 255:red, 74; green, 144; blue, 226 }  ,draw opacity=1 ][fill={rgb, 255:red, 255; green, 255; blue, 255 }  ,fill opacity=1 ] (151.8,219) .. controls (151.8,207.95) and (168.86,199) .. (189.9,199) .. controls (210.94,199) and (228,207.95) .. (228,219) .. controls (228,230.05) and (210.94,239) .. (189.9,239) .. controls (168.86,239) and (151.8,230.05) .. (151.8,219) -- cycle ;
		\draw  [color={rgb, 255:red, 65; green, 117; blue, 5 }  ,draw opacity=1 ][fill={rgb, 255:red, 255; green, 255; blue, 255 }  ,fill opacity=1 ] (355.14,215.5) .. controls (355.14,205.28) and (372.2,197) .. (393.24,197) .. controls (414.28,197) and (431.34,205.28) .. (431.34,215.5) .. controls (431.34,225.72) and (414.28,234) .. (393.24,234) .. controls (372.2,234) and (355.14,225.72) .. (355.14,215.5) -- cycle ;
		\draw [color={rgb, 255:red, 74; green, 144; blue, 226 }  ,draw opacity=1 ] [dash pattern={on 4.5pt off 4.5pt}]  (424.8,119) -- (451.8,119) ;
		\draw [color={rgb, 255:red, 65; green, 117; blue, 5 }  ,draw opacity=1 ] [dash pattern={on 4.5pt off 4.5pt}]  (423.8,149) -- (443.8,149) -- (450.8,149) ;
		\draw  [color={rgb, 255:red, 74; green, 74; blue, 74 }  ,draw opacity=1 ][dash pattern={on 0.84pt off 2.51pt}] (120,57) -- (590.8,57) -- (590.8,288) -- (120,288) -- cycle ;
		\draw (277,112) node [anchor=north west][inner sep=0.75pt]   [align=left] {$\displaystyle \vert\psi\rangle $};
		\draw (169,248) node [anchor=north west][inner sep=0.75pt]   [align=left] {$\displaystyle Alice$};
		\draw (376,244) node [anchor=north west][inner sep=0.75pt]   [align=left] {$\displaystyle Bob$};
		\draw (445,140) node [anchor=north west][inner sep=0.75pt]   [align=left] {\textit{{\scriptsize \textbf{{\fontfamily{pcr}\selectfont Sending Odd state}}}}};
		\draw (444,112) node [anchor=north west][inner sep=0.79pt]   [align=left] {\textit{{\scriptsize \textbf{{\fontfamily{pcr}\selectfont Sending Even state}}}}};
		\draw (216,57) node [anchor=north west][inner sep=0.75pt]   [align=left] {\begin{minipage}[lt]{100.94pt}\setlength\topsep{0pt}
				\begin{center}
					{\fontfamily{ptm}\selectfont \textit{Grauber Coherent State}}\\{\fontfamily{ptm}\selectfont \textit{as entangled state}}
				\end{center}
		\end{minipage}};
		\draw (157,208) node [anchor=north west][inner sep=0.75pt]   [align=left] {\textit{{\small \textbf{{\fontfamily{pcr}\selectfont \textcolor[rgb]{0.29,0.29,0.29}{$\vert Even\rangle$}}}}}};
		\draw (363,204) node [anchor=north west][inner sep=0.75pt]   [align=left] {{\fontfamily{pcr}\selectfont \textbf{\textit{\textcolor[rgb]{0.29,0.29,0.29}{$\vert Odd\rangle$}}}}};
	\end{tikzpicture}
	\caption{BQT using multipartite coherent states $\vert \psi\rangle$, Alice and Bob can communicate their states $\vert Even\rangle$ and $\vert Odd\rangle$ at the same time, At the end, Both sender and the receiver are given a state that has some decoherence.}\label{Fig2}
\end{figure}

\begin{figure}[hbtp]
	{{\begin{minipage}[b]{1.0\linewidth}
				\centering
				\includegraphics[scale=0.3]{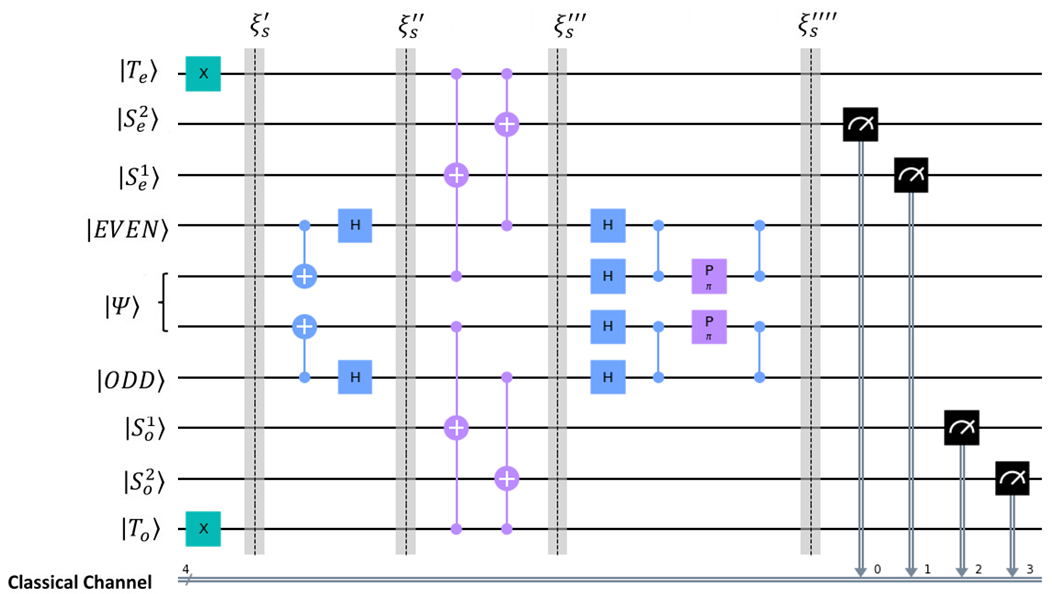} \vfill 
	\end{minipage}}}
	\caption{The implementation of BQT with two independent triggers qubits using Qiskit. The X-gates are applied to set the first and last qubits to the state $\left|1\right\rangle$. Then, we perform controlled-NOT (CNOT) gates between two pairs of qubits, these CNOT gates create entanglement between the qubits. The application of controlled-controlled-NOT (CCNOT) gates between three qubits are used to teleport the state of one qubit to another qubit, and the entangled state of the entire system is generated by putting Hadamard gates (H), Controlled-Z gates, and phase shift gates, the values of two variables, $m$ and overlap $p$, are adjusted to achieve a high fidelity accuracy.}\label{Fig3}
\end{figure}

In this protocol that employs the multipartite coherent state (\ref{eq2}) as a quantum channel, Alice and Bob, as two legitimate users, want to bidirectionally teleport the even and odd coherent states (\ref{Eq10}) to each other, as shown schematically in Fig.(\ref{Fig2}). In this respect, ten qubits are used to implement this BQT scheme, as clearly explained in the quantum circuit (see Fig.\ref{Fig3}). Two individuals Even and Odd who each own a qubit and desire to share a state across a quantum channel are involved in the protocol. It is essential to mention some important steps applied in the implementation of the presented BQT protocol. Firstly, the even and odd coherent states (\ref{Eq10}), in the Bloch sphere representation, are expressed as follows
\begin{align}
&\rho_{Even}^{(e)}=\frac{1}{2}\left(\openone_{2}+\sum_{j=x, y, z} e_{j} \hat{\sigma}_{j}^{(e)}\right),\hspace{1cm}\rho_{Odd}^{(o)}=\frac{1}{2}\left(\openone_{2}+\sum_{j=x, y, z} o_{j} \hat{\tau}_{j}^{(o)}\right),\label{eq20}
\end{align}
where $\openone_{2}$ is a $2\times2$ identity operator, $e_{j}=\operatorname{Tr}(\sigma_{j}\rho_{Even}^{(e)})$ and $o_{j}=\operatorname{Tr}(\tau_{j} \rho_{Odd}^{(o)})$ are the correlation matrix elements, $\hat{\sigma}_{j}^{(e)}$ and $\hat{\tau}_{j}^{(o)}$ denotes the Pauli matrices in the Even and Odd computational basis, respectively. In fact, these states (\ref{Eq10}) are expected to play an important role in interferometric detection of gravitational waves by lowering the ideal intensity of the input laser \cite{Ansari1994}. Furthermore, when performing tests to verify quantum information technology, the size of the coherent components $\eta$ is of utmost importance \cite{An2003,Podoshvedov2012,An2004}.

In the BQT scheme, each user has two distinct types of qubits in order that Alice and Bob can exchange even and odd states with each other, i.e. the trigger qubits that are initially specified to be
\begin{equation}
	\begin{aligned}
		& \left|T_{e}\right\rangle=\cos \left(\frac{\vartheta_{e}}{2}\right)\left|\eta_{e}\right\rangle+\sin \left(\frac{\vartheta_{e}}{2}\right)\left|-\eta_{e}\right\rangle, \\
		& \left|T_{o}\right\rangle=\cos \left(\frac{\vartheta_{o}}{2}\right)\left|\eta_{o}\right\rangle+\sin \left(\frac{\vartheta_{o}}{2}\right)\left|-\eta_{o}\right\rangle,
	\end{aligned}
\end{equation}
with $\vartheta_{i}\in\left[0,\pi\right]$ for ($i=e,o$). These can be represented in the Bloch vector expansion as
\begin{align}\label{eq1} 
	\rho_{T_{e}}=\frac{1}{2}\left(\openone_{2}+\sum_{j=x, y, z} t_{j} \hat{\sigma}_{j}^{(e)}\right),\hspace{1cm}\rho_{T_{o}}=\frac{1}{2}\left(\openone_{2}+\sum_{j=x, y, z} g_{j} \hat{\tau}_{j}^{(o)}\right),
\end{align}
whereas in Eq.(\ref{eq1}), $t_{j}$ and $g_{j}$ are the Bloch vector elements of both qubits Even's and Odd's trigger and those elements are calculated using the formula $e_{j}=\operatorname{Tr}(\sigma_{j}\rho_{T_{e}})$ and $g_{j}=\operatorname{Tr}(\tau_{j}\rho_{T_{o}})$. Besides, two storage qubits are used in BQT, one kept by the sender and the other by the recipient. The two parties exchange and store quantum information using these qubits; $\vert S^{1}_e\rangle$, $\vert S^{2}_e\rangle$ for the even direction and $\vert S^{1}_o\rangle$, $\vert S^{2}_o\rangle$ for the odd direction (see Fig.(\ref{Fig3})).

To perform this protocol, Alice and Bob initially share a state of the form (\ref{rhoij}) and we follow the Kiktenko steps reported in \cite{Kiktenko2016}:\par

\textbf{Step 1:} We start by setting the users quantum state as
\begin{align*}
	\vert\xi_s\rangle = \vert T_{e}, S^{1}_{e},S^{2}_{e}, Even\rangle\vert T_{o}, S^{1} _{o},S^2_{o},Odd\rangle \otimes\vert \Psi\rangle
\end{align*}

\textbf{Step 2:} Applying $X$-gates to the first and final qubits of Triggers qubits, $\vert Even\rangle$ and $\vert Odd\rangle$, sets the two qubits to be shared.

\textbf{Step 3:} We apply two controlled-not (CNOT) gates to the qubits $\vert Even\rangle$ and $\vert Odd\rangle$ and the entangled state $\vert \Psi \rangle$, means that if $\vert Even\rangle$ is in state $|1\rangle$, then the entangled state $\vert \Psi \rangle$ is flipped. Similary for the entangled state $\vert\Psi\rangle$ and $\vert Odd\rangle$ as the control qubit. The equations for these operations can be written as
\begin{align*}
	\vert\xi_s\rangle'=CNOT(\vert T_{e},S^{1}_{e},S^{2}_{e},Even\rangle\vert T_{o}, S^{1}_{o},S^{2}_{o}, Odd\rangle\otimes\vert\Psi\rangle),
\end{align*}
and then applies a Hadamard gate to both qubits $\vert Even\rangle$ and $\vert Odd\rangle$ to prepare these qubits in entangled superposition state. QT can be accomplished in this entangled state
\begin{align}
	\vert\xi_s\rangle''=H&\left[CNOT\left(\vert T_e, S^1 _e,S^2 _e, Even\rangle \vert T_o, S^1 _o,S^2 _o, Odd\rangle \otimes\vert\Psi\rangle\right)\right].
\end{align}
\textbf{Step 4:}  Both Triggers qubits $\vert T_{e,o}\rangle$, the entangled qubit $\vert \Psi\rangle$ and storage qubits $\vert S^i _j\rangle$ with $i = 1,2$, $j= e,o$, are all subject to the two parties perform two Controlled-CNOT $CCNOT$ qubits on them. The entangled qubit $\vert \Psi\rangle$ and triggers qubits $\vert T_{eo}\rangle$ are the controlled qubits, and $\vert S^i _j\rangle$ are the target qubits. we express that as follow:
\begin{align}
	\vert\xi_s\rangle'''=CCNOT&\left[H\left\lbrace CNOT\left(\vert T_{e}, S^1 _{e},S^2_{e}, Even\rangle \vert T_{o}, S^1_{o},S^2_{o}, Odd\rangle \otimes\vert\Psi\rangle\right)\right\rbrace\right].
\end{align}
\textbf{Step 5:} The two parties apply two Hadamard gates $H$ and two Controlled-Z gate on $\vert Even\rangle$ and $\vert Odd\rangle$, desired to send, and they use Phase shift gate $CP$ on the entangled state of the entire system, then use again a Controlled-Z on $\vert Even\rangle$, $\vert Odd\rangle$ and the entangled qubit $\vert \Psi\rangle$. This phase of the protocol describes the actual teleportation of information
{\small
	\begin{align}
		\vert\xi_s\rangle''''=& CZ[CP[CZ\left[H\left(CCNOT\left\lbrace H\left(CNOT \left\lbrace\vert T_e, S^1 _e,S^2 _e, Even\rangle \vert T_{o}, S^1_{o},S^2_{o},Odd\rangle \otimes\vert\Psi\rangle\right\rbrace\right)\right\rbrace\right)\right]]].
	\end{align}}

After following all the phases above, and perform measurements on the four storage qubits. The final teleported states are giving by
\begin{align}\label{101}
	&\rho^{(e)}_{out } = P_{o} \overline{P}_{e} \rho_{Odd}^{(\text {o) }}+\left(1-P_{o} \overline{P}_{e}\right) \rho_{1}, \notag\\&
	\rho_{out}^{(o)}= P_{e} \overline{P}_{o} \rho_{Even}^{\text {(e})}+\left(1-P_{e} \overline{P}_{o}\right) \rho_{1}, 
\end{align}
with $\overline{P}_{i}=1- P_{i}$, $P_{i}= Tr(\rho_{T_i}\rho_j)$ (with $i=e,o$ and $j=Even,Odd$) and $\rho_{1}$ denotes the reduced state of the quantum channel (\ref{rhoij}), written as 
\begin{equation}
\rho_{1}={\rm Tr}_{2}\left[\rho_{12}\right]=\frac{3+p^2 +\left(1+3p^2\right)p^{n-2}\cos(m\pi)}{4\left(1+p^{n}\cos(m\pi)\right)} \begin{pmatrix}
	1 & 0 \\
	0 & 1 
\end{pmatrix}.
\end{equation}
\textbf{Step 6:} The sender applies two Pauli gates (X,Z) to the qubit $\vert Even\rangle$ depending on the outcomes of the measurements, and receiver applies two Pauli gates to his qubit $\vert Odd\rangle$ based on the classical information he obtained from sender's measurements, and vice versa. The Pauli gates that Even and Odd use are determined by the values of the classical bits retrieved from the measurements.
\section{Optimal fidelity in the proposed BQT scheme}\label{Sec3}
The fidelity of quantum teleportation refers to how closely the transported state matches the original state. In other words, it shows us how much information is retained throughout the transfer procedure. Fidelity is commonly stated as a number between $0$ and $1$, with $1$ representing full fidelity. Many factors can influence the reliability of a bidirectional quantum teleportation for coherent states. They include the degree of entanglement between the two participants, the quality of the quantum channel used to transfer the state, and the measurement and reconstruction process's efficiency \cite{Wang2019}. The fidelity is a crucial quantity for measuring the overlap between the density operator $\rho_{\text {out }}$ for the teleported state $\left|\psi_{\text {in }}\right\rangle$ and the quantum state into that is going to be teleported, i.e.,
\begin{equation}
	\mathcal{F}\left(\eta_{e}, \eta_{o}\right)=\left\langle\psi_{\text {in }}\left|\rho_{\text {out }}\right| \psi_{\text {in }}\right\rangle=\left(Tr\left[ \sqrt{\sqrt{\rho_{in}}\rho_{out} \sqrt{\rho_{in}}}\right]\right)^2
\end{equation}
The following provides the fidelity of the teleportation between the two users $\mathcal{F}^{A \rightarrow B}$ and $\mathcal{F}^{B \rightarrow A}$:
\begin{align}
	&\mathcal{F}^{A \rightarrow B}=\left(Tr\left[\sqrt{\sqrt{\rho_{even}}\rho_{out}^{(e)} \sqrt{\rho_{even}}}\right]\right)^{2},\hspace{1cm} \mathcal{F}^{B \rightarrow A}=\left( Tr\left[\sqrt{\sqrt{\rho_{odd}}\rho_{out}^{(o)} \sqrt{\rho_{odd}}}\right]\right)^{2}.
\end{align}
Using Eq.(\ref{eq1}) and Eq.(\ref{eq9}) at an initial state setting, we are able to determine the fidelity of the teleported state for both directions. Thus, the final expression for the fidelity from Alice to Bob is obtained as
\begin{align}
	\mathcal{F}^{A \rightarrow B}= \frac{1}{8}\left(1+p\right)^{2}\left(1+\sin\vartheta_{e}\right)&\left[1+p+\left(1-p\right)\sin\vartheta_{o}\right]+  \frac{\Lambda\left(p,n,m\right)(p^{2}-1)}{32\left[1+p^n \cos\left(m\pi\right)\right]}\left[\left(1+p\right)\sin\vartheta_{o}-p-3+ \right. \notag\\&\left. \left(1+p\right)^{2}\sin\vartheta_{e}\left(\left(p-1\right) \sin\vartheta_{o}-1-p\right)\right],
\end{align}
where the function $\Lambda\left(p,n,m\right)$ is given by
\begin{equation}
\Lambda\left(p,n,m\right)= 3+p^2+\left(3p^{2}+1\right) p^{n-2} \cos(m\pi).
\end{equation}
Likewise, the analytical expression of the fidelity from Bob to Alice is given by
{\footnotesize 
\begin{align}
	\mathcal{F}^{B\rightarrow A} = \frac{p-1}{32 p^2\left(1+p^{n}\cos(m\pi)\right)}&\left[\left( p^2\left(3 + p^2\right)+p^{n}\cos(m\pi)\left(1+3p^2\right)\right)\right.\notag\\
	&\left.  \left(p^2+2p-3+\left(1+p\right)^2 \sin\vartheta_{e}-\left(p^{2}-1\right)\left(1+\sin\vartheta_{e}\right) \sin\vartheta_{o}\right)\right]. 
\end{align}}

\begin{widetext}

\begin{figure}[hbtp]
{{\begin{minipage}[b]{.24\linewidth}
\centering
\includegraphics[scale=0.26]{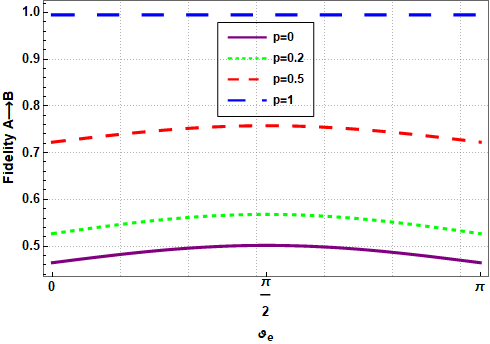} \vfill $\left(a\right)$: $m=0,n=3, \vartheta_{o}=0$.
\end{minipage}\hfill
\begin{minipage}[b]{.24\linewidth}
\centering
\includegraphics[scale=0.26]{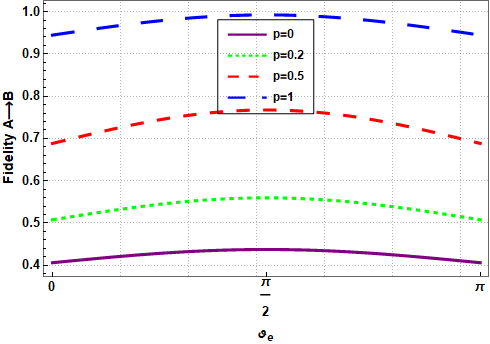} \vfill  $\left(b\right)$: $m=1,n=25,\vartheta_{o}=\pi$.
\end{minipage}\hfill
\begin{minipage}[b]{.24\linewidth}
\centering
\includegraphics[scale=0.26]{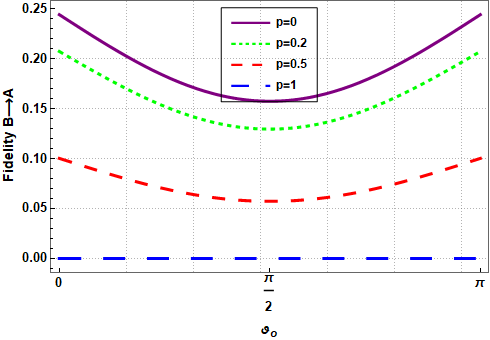} \vfill $\left(c\right)$: $m=0,n=3,\vartheta_{e}=0$.
\end{minipage}\hfill
\begin{minipage}[b]{.24\linewidth}
\centering
\includegraphics[scale=0.26]{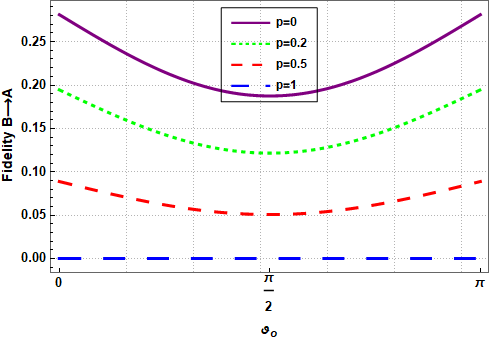} \vfill  $\left(d\right)$: $m=1,n=25,\vartheta_{e}=\pi$.
\end{minipage}}}
\caption{The fidelity of teleportation from Alice to Bob (Panel.($a$) and Panel.($b$)) and from Bob to Alice (Panel.($c$) and Panel.($d$)) as a function of E for different values of the overlap parameter $p$. The comparison is conducted for four different scenarios, each with different values for $m=0$, $m=1$ and $n=3$, $n=25$. The fidelity values are described for each scenario as follows: (a) fidelity from Alice to Bob with $\vartheta_{o}=0$, (b) fidelity from Alice to Bob with $\vartheta_{o}=\pi$, (c) fidelity from Bob to Alice with $\vartheta_{e}=0$, and (d) fidelity from Bob to Alice with $\vartheta_{e}=\pi$.}\label{Fig4}
\end{figure}
\end{widetext}
To show the importance of the choice of the "preshared entanglement state" to work with as a quantum resource on the teleportation efficiency, we depict in the Fig.(\ref{Fig4}) the variation of the fidelity of the BQT with respect to the angle of both sender and receiver teleported coherent states $\nu_{e,o}$ for various value of the overlap $p$, the number of partitions $n$ and the parameter $m$.\par 

To better appreciate this behavior, it is important to consider the limit cases for symmetric ($m$ even) and antisymmetric ($m$ odd) states.  As shown in Fig.(\ref{Fig4}$(a)$), we observed that the fidelity increases as the coherent state overlapping $p$ increases. We also noted that the minimum number of partitions $n=3$ is required for this behavior to occur. This finding indicates that for the case of symmetric state (i.e.,$m=0$), the fidelity reaches its maximum value if the shared resource is a separable state (i.e.,$p\longrightarrow1$). Furthermore, for the antisymmetric state $m=1$ as represented in Fig.(\ref{Fig4}$(b)$), we noticed that the fidelity increases and the curves to be maximum when $p\longrightarrow1$, and in this case the shared resource state is reduced to a bipartite state of Werner-type which is maximally entangled. This maximum value occurs for a maximum number of partitions, e.g. with $n=25$. Besides, when the shared resource is in a maximally intricate state (in particular a GHZ-state; i.e., $m=0$ and $p=0$), the maximum fidelity is achieved during teleportation from an odd coherent state to an even coherent state (see Fig.(\ref{Fig4}$(c)$)). Additionally, Fig.(\ref{Fig4}$(d)$) shows that the fidelity is maximized when the shared state between Alice and Bob is a GHZ maximally entangled state. It corresponds to the antisymmetric state m=1 and the overlap p=0 with a maximum number of partitions $n=25$. These results provide important insights into the optimal conditions for achieving high-fidelity quantum teleportation between even and odd coherent states.\par
 
These findings are critical in understanding the behavior of the fidelity of our quantum system and can be utilized to optimize the performance of the BQT protocol. When we specifically focus on the case where $n=3$, and the overlap parameter $p$ is set to $1$, we observe a perfect fidelity with a value of $1$. This implies that the teleportation process is flawless, and the teleported state is an exact replica of the original state. However, as we vary the overlap parameter $p$ for different values of parameters $m$ and the number of partitions $n$, we notice that the $p=0.2$ and $p=0.5$ exhibit similar trends and shapes with different values for fidelity. Furthermore, our analysis reveals that the fidelity of quantum teleportation tends to improve as the number of qubits $n$ increases, particularly at $n=25$ where we observe the interference of other parameters. $\mathcal{F}^{B\rightarrow A}$ at $p=1$ exhibits a slight curve, indicating that the quantum system is approaching the limits of its performance in terms of fidelity.
\section{Estimation degree of a teleported state}\label{Sec4}
\paragraph{\bf Quantum Fisher Information:} Unlike quantum state teleportation in which their credibility is gauged by teleportation fidelity, the information transmission conveyed by a physical parameter is generally measured using QFI \cite{Paris2009}. In other words, the amount of quantum information is simply the maximum QFI in a measurement of the quantum state, optimized over all possible measurements. Moreover, the ultimate possible precision of the estimated parameter is provided by the inverse of the QFI \cite{SlaouiDrissi2022,ElBakraoui2022}. This provides the sensitivity of a given quantum state with respect to changes in the relevant encoded parameter. Further, it has been shown that by estimating the relevant parameters encoded in the teleported state, with the efficiency quantified by means of QFI, the credibility of the QT protocol could be assessed \cite{Anouz2019,Anouz2020}. Based on the symmetric logarithmic derivative operator $\mathcal{L}_{\xi}$ associated with parameter $\xi$, the corresponding QFI is expressed as $\mathrm{F}_{\xi}=\operatorname{Tr}\left[\left(\partial_{\xi} \rho_{\xi}\right) \mathcal{L}_{\xi}\right]$, where $\partial_{\xi} \rho_{\xi}=\left(\mathcal{L}_{\xi} \rho_{\xi}+\rho_{\xi} \mathcal{L}_{\xi}\right)/2$ and $\partial_{\xi}=\partial/\partial\xi$. In order to derive the analytical expression for the QFI, we write the density operator $\rho_{\xi}$ in its spectral decomposition as $\rho_{\xi}=\sum_{k=1}^{d} \lambda_{k}|\psi_{k}\rangle\langle\psi_{k}|$, where $\lambda_{k}$ and $|\psi_{k}\rangle$ are the eigenvalues and eigenvectors of the matrix $\rho_{\xi}$, then we find that QFI can be reformulated as
\begin{align}
\mathrm{F}_{\xi} =\sum_{k=1}^{d} \frac{\left(\partial_{\xi} \lambda_{k}\right)^{2}}{\lambda_{k}}+\sum_{k=1}^{d} \lambda_{k} \mathrm{F}_{\xi,k} -\sum_{k \neq l} \frac{8 \lambda_{k} \lambda_{l}}{\lambda_{k}+\lambda_{l}}\left|\left\langle\psi_{k} \mid \partial_{\xi} \psi_{l}\right\rangle\right|^{2},
\end{align}
where $\lambda_{k}\neq0$ and $\lambda_{k}+\lambda_{l}\neq0$ with $\mathrm{F}_{\xi,k}=4\left[\left\langle\partial_{\xi}\psi_{k}\mid\partial_{\xi} \psi_{k}\right\rangle-\mid\left\langle\psi_{k}\mid\partial_{\xi} \psi_{k}\right\rangle\mid^{2}\right]$ being the QFI for the pure state $\left|\psi_{k}\right\rangle$. The purpose of this part is to estimate the phase parameters encoded in the teleported states by means of the QFI in both directions; from Alice to Bob and from Bob to Alice. This requires finding the QFI in both teleported even and odd coherent states (eq.(\ref{Eq10})). For single-qubit states like our teleported states, Zhong et al.\cite{Zhong2013} derived a simple and explicit formula for QFI as
\begin{equation}\label{eq34}
	\mathrm{F}_{\xi}\left(\rho_{\xi} \right) =\left\{\begin{array}{cc}
		\left|\partial_{\xi} \vec{r}\right|^{2}+\frac{(\vec{r} \cdot \partial_{\xi} \vec{r})^{2}}{1-|\vec{r}|^{2}} & \text { if }|\vec{r}|<1, \\
		\left|\partial_{\xi} \vec{r}\right|^{2} & \text { if }|\vec{r}|=1,
	\end{array}\right.
\end{equation}
with $\vec{r}=\left(r_{x},r_{y},r_{z}\right)^{T}$ is the real Bloch vector of $\rho_{\xi}$. Indeed, QFI can also be interpreted as a measure of quantum statistical speed; these two notions quantifying the sensitivity of an initial state to the variations of the estimated parameter and the higher this sensitivity is, the more the estimated parameter can be approximated with precision and efficiency. In turn, for parametric evolutions of classical probability distributions, each statistical distance measure is reflected by a statistical speed. This can be evaluated by taking the derivative of the distance, i.e. the variation of the distance caused by a small change. By maximizing the classical statistical speed over all the quantum measurements, we then obtain the quantum statistical speed \cite{Jeffreys1946,Gessner2018}. In the following, we analyze another type of quantum statistical speed named Hilbert-Schmidt Speed, and explore how it can be related to QFI and fidelity in our BQT protocol.
\paragraph{\bf Hilbert-Schmidt Speed:} Let's start by recalling the family of distance measures such as
\begin{equation}
	|d_{\alpha}(p,q)|^{\alpha}=\frac{1}{2}\sum_{x}|p_{x}-q_{x}|^{\alpha},\label{35}
\end{equation}
where $q=\{q_{x}\}_{x}$ and $p=\{p_{x}\}_{x}$ depend on the parameter $\xi$ and represent the probability distributions, with  $\alpha\geq 1$. To derive the associated statistical speed from any statistical distance, we need to quantify the distance between infinitesimally close distributions. We first parameterize the probability distribution $p(\xi)$ and extend it to first order in $\xi$ at $\xi_{0}$ (Taylor expansion at $\xi_{0}$), i.e.,
\begin{equation}
	p_{x}(\xi_{0}+\xi)=p_{x}(\xi_{0})+\frac{\partial p_{x}(\xi)}{\partial\theta}|_{\xi=\xi_{0}}\xi+{\cal O}(\xi^{2}).\label{36}
\end{equation}
Using this expansion and substituting expression (\ref{36}) into equation (\ref{35}), the distance between $p(\xi_{0})$ and $p(\xi_{0}+\xi)$ is evaluated as
\begin{equation}\label{37}
	d_{\alpha}(p(\xi_{0}+\xi),p(\xi_{0}))=\left(\frac{1}{2}\sum_{x}|p'_{x}(\xi_{0})|^{\alpha}\right)^{1/\alpha}\xi+{\cal O}(\xi^{2}),
\end{equation}
with $p'_{x}(\xi_{0})=\frac{\partial p_{x}(\xi)}{\partial\xi}|_{\xi=\xi_{0}}$. Using the above classical distance (\ref{37}), the classical statistical speed is given by
\begin{equation} \label{classical speed}
	s_{\alpha}[p(\xi_{0})]=\frac{d}{d\xi}\left[d_{\alpha}(p(\xi_{0}+\xi),p(\xi))\right]=\left(\frac{1}{2}\sum_{x}|p'_{x}(\xi_{0})|^{\alpha}\right)^{1/\alpha}
\end{equation}
Taking now the quantum states $\rho$ and $\sigma$, we can expand these classical concepts to the quantum case by taking $q_{x}={\rm Tr}\{E_{x}\sigma\}$ and $p_{x}={\rm Tr}\{E_{x}\rho\}$ as the measurement probabilities associated with the positive operator value measure (POVM) satisfying $\sum_{x}E_{x}=\openone$ and $E_{x}\geq0$. In this situation, the quantum distance can be obtained by maximizing the Hellinger distance (\ref{classical speed}) over all possible POVMs, that is,
\begin{equation}
	D_{\alpha}(\rho,\sigma):=\max_{\{E_{x}\}}d_{\alpha}(p,q)=\left(\frac{1}{2}{\rm Tr}|\rho-\sigma|^{\alpha}\right)^{1/\alpha},
\end{equation}
and the statistical quantum speed is given by
\begin{equation}
	S_{\alpha}[\rho_{\xi}]=\max_{\{E_{x}\}}s_{\alpha}[p(\xi)]=\left(\frac{1}{2}{\rm Tr}|\frac{d\rho_{\xi}}{d\xi}|^{\alpha}\right)^{1/\alpha}.\label{40}
\end{equation}
In the special case where $\alpha=2$, the quantum statistical speed is reduced to the Hilbert-Schmidt speed (HSS) and equation (\ref{40}) thus becomes
\begin{equation} \label{HSS Equation}
	{\rm HSS}[\rho_{\xi}]\equiv {\rm HSS}_{\xi}\equiv S_{2}[\rho_{\xi}]=\sqrt{\frac{1}{2}{\rm Tr}|\frac{d\rho_{\xi}}{d\xi}|^{2}}.
\end{equation}
Notably, the HSS is recognized as a powerful figure of merit in refs.\cite{Jahromi2020,Abouelkhir2023,Rangani2021} to improve quantum phase estimation as well as to identify non-Markovianity in an open quantum system consisting of $d$-qubits. More generally, it would seem logical to explore the connections between HSS and QFI since they are two quantum statistical speeds connected to the Hilbert-Schmidt and Bures distances \cite{Braunstein1994,Ozawa2000} respectively.\par

To determine the analytical expression for QFI and HSS of teleported states (\ref{101}) from Eq.(\ref{eq34}), it is necessary to compute the three corresponding Bloch vector components ($e_{j}$ and $o_{j}$) given by Eq.(\ref{eq20}). After a few simplifications, the result is
\begin{align}
&o_{x}=2 P_{e} \overline{P_{o}} + (1-P_{e} \overline{P_{o}}) \frac{3+p^2+(1+3p^2)p^{n-2} \cos(m\pi)}{4(1+p^{n})\cos(m\pi)},\notag\\&o_{y}= -2 (1-P_{e} \overline{P_{o}}) \frac{3+p^2+(1+3p^2)p^{n-2} \cos(m\pi)}{4(1+p^{n})\cos(m\pi)},\notag\\&o_{z}=-P_{o} \overline{P_{e}},\label{eq35}
\end{align}
for the even coherent state, and
\begin{align}
&e_{x}= 2 P_{o} \overline{P_{e}} + (1-P_{o} \overline{P_{e}}) \frac{3+p^2+(1+3p^2)p^{n-2} \cos(m\pi)}{4(1+p^{n})\cos(m\pi)},\notag\\&e_{y}= -2 (1-P_{o} \overline{P_{e}}) \frac{3+p^2+(1+3p^2)p^{n-2} \cos(m\pi)}{4(1+p^{n})\cos(m\pi)},\notag\\&e_{z}= -P_{o} \overline{P_{e}},\label{eq36}
\end{align}
for the odd coherent state with the entries $P_{e,o}$ given by
\begin{align}
P_{e,o}= \frac{1}{2}(1-p)\left[1 - 2\cos\vartheta_{e,o}\sin\vartheta_{e,o}\right].
\end{align}
By inserting equations (\ref{eq35}) and (\ref{eq36}) into equation (\ref{eq34}) with respect to the parameters $\vartheta_{e}$ and $\vartheta_{o}$, the QFI takes the form
\begin{align}
	F_{\vartheta_{e}}^{A\rightarrow B}= & \chi(n,m,p)+\frac{F^{A\rightarrow B}_{1}\left(n,m,p\right)}{F^{A\rightarrow B}_{2}\left(n,m,p\right)},
\end{align}
from Alice to Bob and 
\begin{align*}
	F_{\vartheta_{o}}^{B\rightarrow A}= \Omega(n,m,p)+\frac{F^{B\rightarrow A}_{1}(n,m,p)}{F^{B\rightarrow A}_{2}(n,m,p)},
\end{align*}
from Bob to Alice. Consequently, the analytical expressions of the SHSs, associated to the estimated trigger phases of Alice $\vartheta_{e}$ and Bob $\vartheta_{o}$, are given by
\begin{align*}
	{\rm HSS}_{\vartheta_{e}}^{A\rightarrow B}= \frac{1}{2}\sqrt{F_{\vartheta_{e}}^{A\rightarrow B}},\hspace{1cm}
	{\rm HSS}_{\vartheta_{o}}^{B\rightarrow A}= \frac{1}{2}\sqrt{F_{\vartheta_{o}}^{B\rightarrow A}},
\end{align*}
with the quantities $F^{A\rightarrow B}_{i}\left(n,m,p\right)$ and $F^{B\rightarrow A}_{i}\left(n,m,p\right)$ are
\begin{widetext}
\begin{align}
F^{A\rightarrow B}_{1}\left(n,m,p\right)&=\cos^{2}\vartheta_{e}\left[ \Upsilon\left(n,m,p\right)+ (p-1)^2(1+p)(1+p^n \cos(m\pi))^2(p-1+(1+p) \sin\vartheta_{e})\left(\cos\left(\vartheta_{o}/2\right) -\sin\left(\vartheta_{o}/2\right)\right)^4-\right. \notag\\ &\left. 
(1-p^{2})\Theta(n,m,p) (1-\sin\vartheta_{o})\left(4+4p^{n} \cos(m\pi)-\frac{1}{4} \Theta(n,m,p)\Gamma(n,m,p)\right)\right]^{2}, 
\end{align}
\begin{align}
F^{A\rightarrow B}_{2}\left(n,m,p\right)&=256(1+p^n \cos(m\pi))^4\left[  1+\frac{1}{64}\left( -4(p-1)^{2}\left(p-1+(1+p)\sin\vartheta_{e}\right)^2(\cos(\vartheta_{o}/2)-\sin(\vartheta_{o}/2))^4-\right.\right.  \notag\\&\left. \left.\Lambda(n,m,p)^{2}- \left(4(1-p)(p-1+(1+p)\sin\vartheta_{e})(\sin\vartheta_{o}-1)+ \Lambda(n,m,p)\right)^{2}\right)\right], 
\end{align}
\begin{align}
F^{B\rightarrow A}_{1}(n,m,p)= & (p^{2}-1)^{2} \cos(\vartheta_{o})^{2}\Xi(n,m,p)^{4}\left[4(1+p)(1+p^n cos(m \pi))^2 (\Xi(n,m,p))^2 \mu(n,m,p) \right.\notag\\  
&\left. +((3+p^2)p^{-2}+p^{n-4} (1+3p^{2}) \cos(m\pi))^{2} \left[(1+p)\sin\vartheta_{e} \mu(n,m,p)+ \kappa (n,m,p)\right]+ \right.\notag\\ 
& \left. 4(p^{2}-1)(1-p^{n-2}\cos(m\pi))\left[ \varepsilon(n,m,p)^{2}\Xi(n,m,p)^{2} \mu(n,m,p) +\right.\right.\notag\\ 
& \left.\left. \frac{1}{4}(3+p^{2}+(3p^{n}+p^{n-2}) \cos(m\pi))\left((1+p) \mu(n,m,p)\sin\vartheta_{e}+\kappa(n,m,p)\right)\right]\right]^{2},
\end{align}
\begin{align}
F^{B\rightarrow A}_{2}(n,m,p)= &4096(1+p^{n}\cos(m\pi))^{4}\left[ 1-\Delta(n,m,p)^{2}\varsigma(n,m,p)^{2} -  \frac{1}{4}\tau(n,m,p)^{2}(1-\Delta(n,m,p)\varsigma(n,m,p))^{2} \right.\notag\\
	&\left.-4\left(\Delta(n,m,p)\varsigma(n,m,p)+\frac{1}{4}\tau(n,m,p)\left(1-\Delta(n,m,p) \varsigma(n,m,p)\right)\right)^{2}\right], 
\end{align}
where the functions $\chi,\Gamma,\Theta,\Lambda,\Upsilon,\Omega,\Xi,\Delta,\varepsilon,\varsigma,\kappa,\mu,\tau,$ are given by
\begin{align*}
&\chi\left(n,m,p\right)= \frac{(p^{2}-1)^2\left(p^{2}(2+p^2)+p^n(1+2p^2)\cos(m\pi)\right)^2 \cos^2\vartheta_{e}(-1+\sin\vartheta_{o})^2}{16 p^4\left(1+p^{n} \cos(m\pi)\right)^2},\notag\\&
\Gamma(n,m,p)=3+2 p-p^2+(p^{2}-1)\sin\vartheta_{e}\left[ \sin\vartheta_{o}-1+(p-1)^2 \sin\vartheta_{o}\right], \hspace{0.5cm}\Theta(n,m,p)=3+p^{2}+\left(3+\frac{1}{p^2}\right)p^{n}\cos(m\pi),\notag\\&
\Lambda(n,m,p)=\frac{\left(3+p^2+(3p^2+1)p^{n-2}\cos(m\pi)\right)\left(3+2 p-p^2+(p^{2}-1)\sin\vartheta_{e}\left(\sin\vartheta_{o}-1\right)+(p-1)^2 \sin\vartheta_{o}\right)}{1+p^{n}\cos(m\pi)},\notag\\&	\Omega(n,m,p)=\frac{(p^{2}-1)^2\left(p^2\left(2+p^2\right)+p^n\left(1+2 p^2\right)\cos(m\pi)\right)^{2} \cos(\phi)^2\left(\cos(\vartheta_{e}/2)+\sin(\vartheta_{e}/2)\right)^4}{16p^4\left(1+p^{n}\cos(m\pi)\right)^2},
\end{align*}
\begin{align*} 
\Upsilon(n,m,p)&= \frac{1}{p^2}(p^{2}-1)^{2} (p^2-p^{n}\cos(m\pi))(-1+\sin (\vartheta_o))\notag\\&\left[(1-p)(1+p^{n} \cos(m\pi))(p-1+(1+p)\sin\vartheta_{e})(-1+\sin\vartheta_{o})+\frac{1}{4}\Theta(n,m,p)\Gamma(n,m,p)\right] 
\end{align*}
	\begin{align*}
		\Xi(n,m,p)& =\cos(\vartheta_{e}/2) + \sin(\vartheta_{e}/2),\hspace{2cm}
		\Delta(n,m,p) =\frac{1}{2}(1+p)\left[1+2\cos(\vartheta_{e}/2)\sin(\vartheta_{e}/2)\right],\\
		\varepsilon (n,m,p) &= -(1+p)(1+p^n \cos(m \pi)),\hspace{2cm}
		\varsigma(n,m,p) =1-\frac{1}{2}(1-p)\left[1-2 \cos(\vartheta_{o}/2) \sin(\vartheta_{o}/2)\right]\\
		\kappa(n,m,p) & =(1-p)\left[3+p-(1+p) \sin(\vartheta_{o})\right],\hspace{2cm}
		\mu(n,m,p) =(p-1)\sin(\vartheta_{o})-1-p,\\
		\tau (n,m,p) &= \frac{1}{(1+p^{n}\cos(m\pi))}\left(3+p^2+p^{n-2}(1+3 p^{2}) \cos(m\pi)\right).
	\end{align*}
\begin{figure}[hbtp]
{{\begin{minipage}[b]{.24\linewidth}
\centering
\includegraphics[scale=0.26]{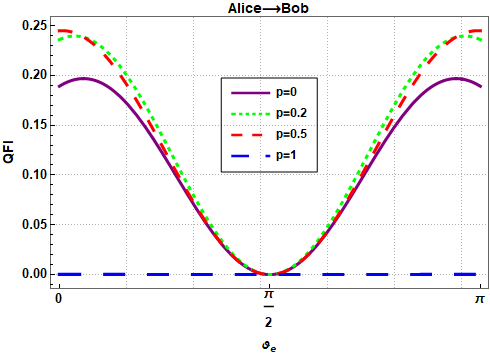} \vfill $\left(a\right)m=0,n=3, \vartheta_{o}=0$
\end{minipage}\hfill
\begin{minipage}[b]{.24\linewidth}
\centering
\includegraphics[scale=0.26]{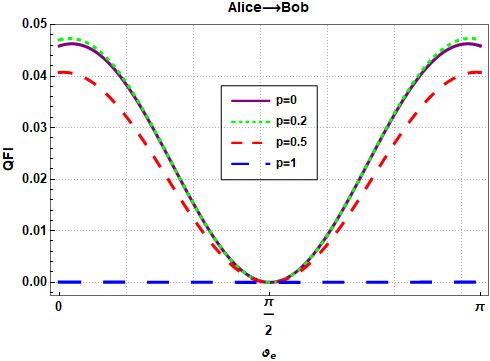} \vfill  $\left(b\right)m=1,n=25,\vartheta_{o}=\pi/6$
\end{minipage}\hfill
\begin{minipage}[b]{.24\linewidth}
\centering
\includegraphics[scale=0.26]{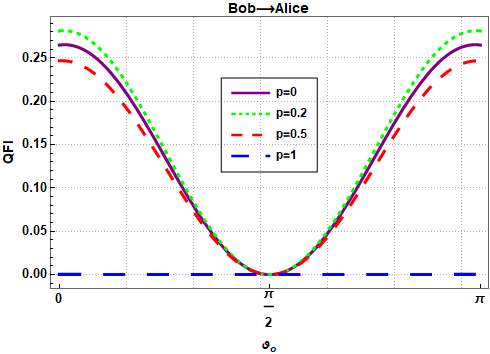} \vfill $\left(c\right)m=0,n=3,\vartheta_{e}=0$
\end{minipage}\hfill
\begin{minipage}[b]{.24\linewidth}
\centering
\includegraphics[scale=0.26]{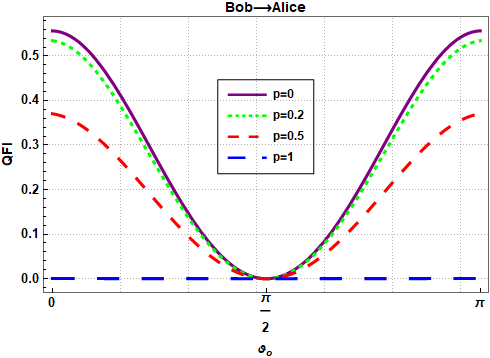} \vfill  $\left(d\right)m=1,n=25,\vartheta_{e}=\pi/6$
\end{minipage}}}\\
	{{\begin{minipage}[b]{.24\linewidth}
			\centering
			\includegraphics[scale=0.30]{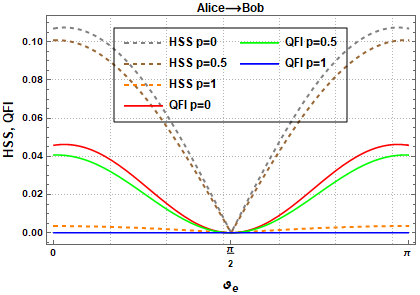} \vfill $\left(e\right)m=0,n=3,\vartheta_{o}=0$
		\end{minipage}\hfill
		\begin{minipage}[b]{.24\linewidth}
			\centering
			\includegraphics[scale=0.30]{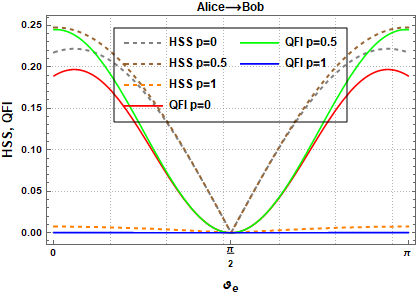} \vfill $\left(f\right)m=1,n=25, \vartheta_{o}=\pi/6$
		\end{minipage}\hfill
		\begin{minipage}[b]{.24\linewidth}
			\centering
			\includegraphics[scale=0.30]{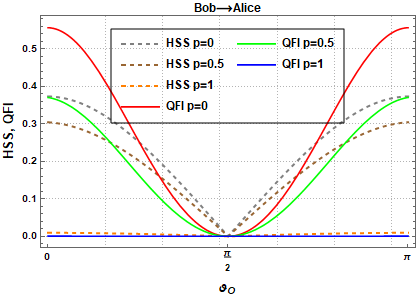} \vfill $\left(g\right)m=0,n=3,\vartheta_{e}=0$
		\end{minipage}\hfill
		\begin{minipage}[b]{.24\linewidth}
			\centering
			\includegraphics[scale=0.30]{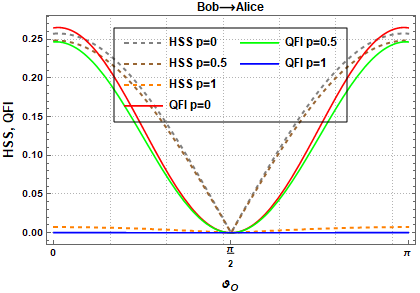} \vfill  $\left(h\right)m=1,n=25,\vartheta_{e}=\pi/6$
\end{minipage}}}
\caption{The top row represents the QFI, same as Fig.(\ref{Fig4}), with different parameters: (a) from Alice to Bob with $m=0,n=3,\vartheta_{o}=0$, (b) from Alice to Bob with $m=1,n=25,\vartheta_{o} =\pi/6$, (c) from Bob to Alice with $m=0,n=3,\vartheta_{e}=0$, and (d) from Bob to Alice with $m=1,n=25,\vartheta_{e}=\pi/6$. The bottom row shows the comparison between QFI and HSS for Alice and Bob's states.}\label{Fig5}
\end{figure}
\end{widetext}
In the top row of Fig.(\ref{Fig5}), we analyzed the behavior of the QFI as a function of the trigger qubits phases $\vartheta_{e,o}$ under the same parameters as in Fig.(\ref{Fig4}). Obviously, the QFI of the teleported states strongly depends on the number of probes $n$, the parameter $m$ and the overlapping $p$ between the coherent states, which specify the pre-shared quantum channel. In the direction from Alice to Bob (Fig.\ref{Fig5} panels (a)-(b)), it is obvious that the QFI of the teleported even coherent state increases with decreasing overlapping parameter $p$. Comparing this behavior with the above results displayed in Fig.(\ref{Fig4}), we conclude that the maximum fidelity corresponds to the minimum variance of the estimated parameter $\nu_{e}$. Additionally, perfect fidelity corresponds to zero QFI, where the variance of the Alice's trigger phase is diverged (see the limiting case $p\longmapsto1$). In the second direction corresponding to Bob towards Alice, and for fixed parameters n and m, the QFI (Fig.\ref{Fig5} panels (c)-(d)) and the teleportation fidelity (Fig.\ref{Fig4} panels (c)-(d)) increase with the decrease of the overlapping $p$. This indicates that the teleportation efficiency reaches to their tops values when the estimation of Bob's trigger phase diverges (i.e. the variance is maximal). More importantly, the behavior of QFI is similar in both directions, but as QFI from Alice to Bob $F_{\vartheta_{e}}^{A\rightarrow B}$ increases, QFI from Bob to Alice $F_{\vartheta_{o}}^{B\rightarrow A}$ decreases, which means that when the information in Alice's trigger hand begins to increase, the information in Bob's trigger hand decreases.\par
To compare quantitatively the QFI and HSS associated with the parameters of the trigger phases, we plotted in the bottom row of Fig.(\ref{Fig5}) their behaviors under the same conditions. From Fig.\ref{Fig5}(panel $(e)$ to panel $(h)$), it is obvious that the minimum and maximum values of the HSS and QFI coincide perfectly and both exhibit oscillatory features. These results provide a qualitative indication that HSS can be employed to pinpoint the precise moments when the estimated trigger qubit phases reach their best estimates. HSS therefore performs the same role as QFI in improving the efficiency of our BQT protocol in both directions.\par

This BQT protocol emerges as an exceptionally advantageous approach, particularly suitable for scenarios necessitating simultaneous data exchange among distributed participants. In our present study, we have deliberately opted for the multipartite state of the entire system as our chosen quantum channel. The utilization of multipartite Glauber coherent states as the quantum channel for BQT brings forth numerous advantages that surpass those associated with other types of entangled states such as GHZ states, cluster states, or Bell states \cite{Zhang2023,Ban2022,Kim2023}. A punching characteristic of coherent states is their resistance to specific types of noise and decoherence, which separates them from far more entangled systems. This fundamental robustness proves to be crucial, particularly in practical implementations where environmental disturbances provide difficult challenges. The feasibility and technical feasibility linked to the creation and control of coherent states endorse their favorability, especially in experimental setups where establishing and sustaining complex entanglement structures like GHZ \cite{Harraz2023,Verma2020} or cluster states \cite{Sang2016,Zhou2019} could present substantial obstacles. Furthermore, the versatility of multipartite Glauber coherent states extends beyond teleportation, making them valuable not only for this purpose but also as potential central components within quantum networks due to their adaptability across a diverse range of quantum communication tasks. When compared to multi-hop states \cite{Zhang2023}, multipartite Glauber coherent states show their advantages as a quantum channel, like in our case. Notably, when considering the impact of noise and diminution of signals during quantum transmission, the robust property of Multipartite Glauber Coherent States becomes even more important. The selection of a right quantum channel shows up as a critical determinant in the field of quantum communication. While the benefits of multi-hop states are undisputed, the unique advantages of Multipartite Glauber Coherent States, such as their reliability, simple encoding, integration with classical communication, decoherence reduction, and efficient resource utilization, make them a more interesting option for various quantum communication paradigms. Coherent states advantages and unique properties position them not only as a theoretical construct, but as a possible trigger in shifting the quantum communication domain into pragmatic practical uses.\par

When examining the work \cite{Zhang2023}, we can perceive the advantages of our BQT protocol. In ref.\cite{Zhang2023}, the teleportation process is highly reliable, with a one hundred percent success rate in an ideal environment. This stability is a significant advantage for secure and effective long-distance quantum communication. We discovered that in the given protocol, the authors have chosen the maximally entangled state (GHZ-type) alongside Bell states as a quantum resource, which has improved the efficiency of fidelity. The protocol can achieve a maximum fidelity of $1$ when certain conditions are satisfied. Additionally, it demonstrates efficiency in terms of achieving perfect teleportation fidelity under specific situations and maintains resilience with regard to different measurement outputs. In our proposed protocol, we adopt a distinct point of view. We treat the whole system state (including the GHZ, ground and W type states) as a quantum resource. During our fidelity calculations, we have demonstrated that we too may obtain maximum fidelity (as in ref.\cite{Zhang2023}) by correctly modifying the parameter values we vary ($m$ for symmetric and antisymmetric states, $n$ for the number of partitions, and $p$ for overlap). This proves that achieving maximum BQT efficiency is consistently accomplished through the utilization of maximally entangled resource states (i.e., GHZ or W states), and this will be attained in our BQT protocol by optimizing the relevant parameters (see Fig.(\ref{Fig4})).
\begin{table}
	\centering
	\begin{tabular}{|p{2cm}||p{2.2cm}|p{5cm}|p{5cm}| }
		\hline
		\multicolumn{4}{|c|}{Contrast between the proposed BQT protocol and prior studies} \\
		\hline
		Protocol & Type of Protocol & Quantum channel & Teleported state \\
		\hline
		Ref.\cite{Zhang2023} & BQT & Multi-hop (N-GHZ state and two Bell states) & Single state  \\
		\hline
		Ref.\cite{Kim2023} & SQT & Single Bell state & Single state \\
		\hline
		Ref.\cite{Harraz2023} & SQT & Tripartite entangled GHZ state and one Bell state  & Single state \\
		\hline
		Ref.\cite{Sang2016}& BQT & Five cluster qubits & Alice with arbitrary two-qubit state and Bob with unknown state  \\
		\hline
		Ref.\cite{Verma2020} & BQT & Two GHZ-States & single-qubit states \\
		\hline
		Ref.\cite{Zhou2019} & BQT & Six-Qubit Cluster State & single-qubit states \\
		\hline
		Our proposed protocol & BQT & Multi-partite coherent state & even and odd coherent states \\
		\hline
	\end{tabular}
	\caption{Comparative Analysis of Seven BQT and QT Protocols in Various Studies}
\end{table}

\section{Experimental implementation of the proposed BQT}\label{Sec6}
Actually, quantum communication protocols are extremely sensitive, i.e. a small variation in some of their parameters renders the protocol completely useless. Countless protocols have been published over the past few decades with supposed improvements on traditional teleportation, and in recent years, we have had free online access to quantum computers and can implement these protocols, and we have observed that the vast majority do not do what the authors proclaimed. These protocols can be implemented in Quirk \cite{Migdal2022}, Qiskit-IBM-Q \cite{Neha2023}, Quantum Inspire by QuTech \cite{Last2020}, Rigetti Forest \cite{Lamm2018}, Quantum Programming Studio \cite{Mastriani2019}, D-Wave \cite{Negre2020} and more. So, to give impartiality and credibility to the implementation, we test here our proposed BQT protocol on Qiskit.\par

Let's assume that the scheme (\ref{Fig3}) represents the complete transmission network. Here, Alice and Bob are the convenient names for the sender and receiver and they wish to send two qubits to be transmitted between them (even and odd coherent states), and both parties have access to a conventional communication channel. Alice's measures it particle and one half of it entangled state in order to transfer the quantum state, and it sends Bob the measurement result over a conventional communication channel. Bob transfers Alice's particle's quantum state to it own by performing a rectification operation on it own particle in accordance with the measurement result. It's crucial to remember that quantum teleportation is probabilistic, which means it could not always work. The success rate of quantum teleportation can, however, approach $100\%$ with enough tries.\par

In fact, BQT is a sort of QT that allows quantum states to be sent between two parties as well as one way alone. BQT requires that both parties own their own trigger qubits, which are used to store the unknown states that they wish to share with one another. To create these states in advance, certain gates are applied to the trigger qubits. Additionally, a traditional communication channel and an entangled state must be shared by the parties. To transmit the quantum states, both sides perform an entanglement measurement on their trigger qubits and the entangled state. They then communicate the results of this measurement across a classical communication channel. Based on the results of the measurements, both parties perform correction operations on their trigger qubits, thereby transferring the quantum states from one party to the other. Below, we describe possible BQT with two random trigger qubits. Also note that the circuit operate with $10$-qubits \cite{Graham2022}.

\begin{table}[h!]
	\centering
	\begin{tabular}{|p{3.5cm}||p{2.5cm}|p{2.2cm}|}
		\hline
		\multicolumn{3}{|c|}{Qubits in the BQT circuit} \\
		\hline
		Sender and receiver & Qubits in Qiskit & States\\
		\hline
		\multirow{3}{*}
		{Even coherent state} & $\mathrm{qr}[0]$ & $\left|T_{e}\right\rangle$ \\
		& $\mathrm{qr}[1]$ & $\left|S_{e}^{2}\right\rangle$ \\
		& $\mathrm{qr}[2]$ & $\left|S_{e}^{1}\right\rangle$ \\
		& $\mathrm{qr}[3]$ & $|Even\rangle$ \\
		\hline
		Entangled state & $\mathrm{qr}[4], \mathrm{qr}[5]$ & $|\psi\rangle$ \\
		\hline
		& $\mathrm{qr}[6]$ & $|Odd\rangle$ \\
		Odd coherent state & $\mathrm{qr}[7]$ & $\left|S_{o}^{1}\right\rangle$ \\
		& $\mathrm{qr}[8]$ & $\left|S_{o}^{2}\right\rangle$ \\
		& $\mathrm{qr}[9]$ & $\left|T_{o}\right\rangle$ \\
		\hline
	\end{tabular}
	\caption{Description of the qubits in the circuit for both sender and receiver}
\end{table}
The circuit begins by preparing the trigger qubits $\left|T_{e}\right\rangle$ and $\left|T_{o}\right\rangle $ in unknown states using X gates. Then, it creates two entangled pairs using controlled-NOT(CNOT) gates and Hadamard (H) gates on qubits $|Even\rangle, |Odd\rangle$ and  $|\psi\rangle $ that is initialized with the entangled state and they serve as a means of transmitting quantum information, while $|Even\rangle$ and $|Odd\rangle $ both states that we want to send and receive. Both qubits $|Even\rangle$ and $|Odd\rangle$ are controlled qubits, when $|\psi\rangle$ is target qubits. Next, the circuit performs a Controlled-Controlled Not (CCNOT) gate on each pair of trigger qubits $\left|T_{e}\right\rangle , \left|S_{e}^{1}\right\rangle  ,|\psi\rangle,\left|S_{e}^{2}\right\rangle,|Even\rangle $ on Even's side and on qubits $\left|T_{o}\right\rangle , \left|S_{o}^{1}\right\rangle, \left|S_{o}^{1}\right\rangle, |Odd\rangle , |\psi\rangle $ on Bob's side, with $\left|S_{e}^{i}\right\rangle$ and $\left|S_{o}^{i}\right\rangle$ ($i=1,2$ look back at the table) are initialize with the state $|0\rangle$, and one half of the corresponding entangled pair. This entangles the trigger qubits with the entangled pairs and collapses their state into one of four possible outcomes. Applying Hadamard gates to qubits $|Odd\rangle$ and $|\psi\rangle$, Controlled-Z gate between qubits $|Odd\rangle$ and $|\psi\rangle $, a phase shift gate with angle $m\pi$ to qubit $|\psi\rangle $, and Controlled-Z gate between qubits $|Odd\rangle$ and $|\psi\rangle $. Same application was done to Alice's side including the entangled state of the whole system. After that, the circuit performs a entanglement measurement on each pair of trigger qubits and one half of the entangled pair, and sends the measurement results to the other party via classical registers. Finally, the circuit measures the final state of the trigger qubits and stores the results in the classical registers.\par

The application of Controlled-Not and Hadamard procedure are used for qubits $|Even\rangle,|Odd\rangle$ and $|\psi\rangle$. The Controlled-Controlled-Not gates creat kind of a "copy" of $|Even\rangle,|Odd\rangle$ and $|\psi\rangle$ and store them in both $\left| S_{e}^{i}\right\rangle, \left| S_{o}^{i}\right\rangle$ with ($i=1,2$) if both triggers are $\left|T_{e}\right\rangle$ and $\left|T_{o}\right\rangle$ are in the state $|1\rangle$. the procedure is repeated in reverse so we can measure qubits $\left|S_{e}^{i}\right\rangle$, $\left.S_{o}^{i}\right\rangle$ ($i=1,2$) sent to opposite parties. The state sent From $|Even\rangle$ to $|Odd\rangle$ and in reverse are obtained on qubit $|\psi\rangle$. The Hadamard gate puts the qubit in a superposition state, equal probability of being in the $\left|0\right\rangle$ state or the $\left|1\right\rangle$ state. It is commonly used to create superposition and entanglement. The CZ gate is a controlled phase gate that performs a phase flip to the target qubit $|\psi\rangle$ if the control qubit $|Odd\rangle$ is in the state $\left|1\right\rangle$. It entangles the qubits and introduces a controlled phase shift. The phase shift gate introduces a phase shift to the state of the qubit. In this case, the angle of the phase shift is determined by the value of $m$ multiplied by $\pi$.
\begin{figure}[hbtp]
	{{\begin{minipage}[b]{.33\linewidth}
				\centering
				\includegraphics[scale=0.3]{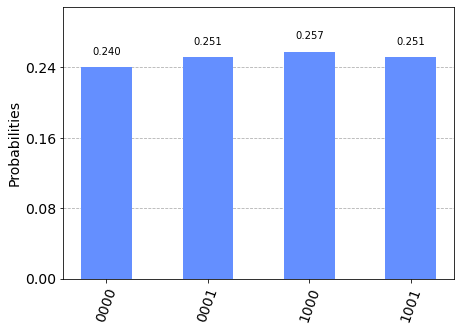} \vfill $\left(a\right) p=0 , m=1$
			\end{minipage}\hfill
			\begin{minipage}[b]{.33\linewidth}
				\centering
				\includegraphics[scale=0.3]{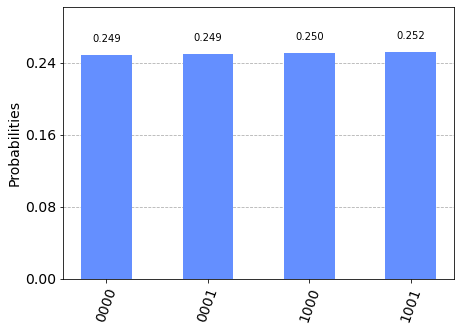} \vfill  $\left(b\right) p=0 , m=1$
	\end{minipage}}}
{{\begin{minipage}[b]{.33\linewidth}
			\centering
			\includegraphics[scale=0.3]{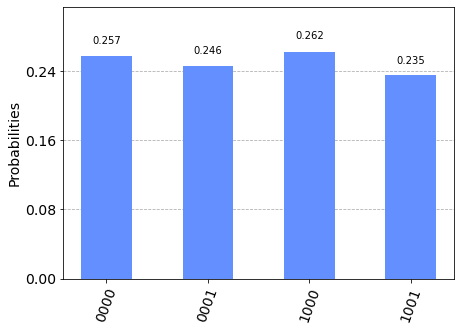} \vfill  $\left(c\right)p=0,m=1$
		\end{minipage}\hfill\\
	\begin{minipage}[b]{.33\linewidth}
		\centering
		\includegraphics[scale=0.3]{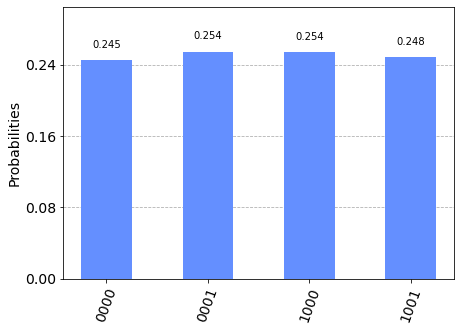} \vfill $\left(d\right)p=1,m=0$
	\end{minipage}\hfill
		\begin{minipage}[b]{.33\linewidth}
			\centering
			\includegraphics[scale=0.3]{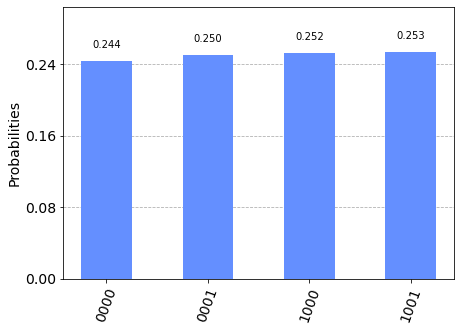} \vfill  $\left(e\right)p=1,m=0$
\end{minipage}\hfill
\begin{minipage}[b]{.33\linewidth}
\centering
\includegraphics[scale=0.3]{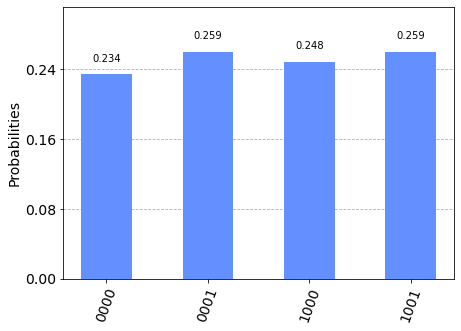} \vfill  $\left(f\right)p=1,m=0$
\end{minipage}}}
	\caption{Histogram shows the output probability (outcomes) obtained by Aer Simulator (Panels ($a$) and ($d$)), Qasm Simulator (Panels ($b$) and ($e$)) with 8192 shots, and Belem Simulator (Panels ($c$) and ($f$)) with fixed values of $m$ and $p$ where the fidelity is theoretically maximum; The top row (panels ($a$)-($b$)-($c$)) shows the results from Alice to Bob when $p=0$ and $m=1$. The bottom row (panels ($d$)-($e$)-($f$)) gives the results obtained from Bob to Alice when $p=1$ and $m=0$.}\label{Fig6}
\end{figure}

The probabilities we obtained for the output states "$0000$", "$0001$", "$1000$", and "$1001$" are the result of simulating the given quantum circuit using the qasm and aer simulators. These probabilities represent the likelihood of measuring the quantum system in the corresponding classical states. In the given code, after applying the gates and measurements, the circuit has four qubits $\left|S_{e}^{i}\right\rangle$, $\left.S_{o}^{i}\right\rangle$ ($i=1,2$) that are measured and mapped to the classical registers. When we run the circuit on a simulator, it generates a set of measurement outcomes based on the probabilities of different classical states. The probabilities are determined by the amplitudes of the quantum states after the measurements. These amplitudes are affected by the gate operations performed on the qubits. The reason we obtained probabilities close to $0.25\%$ for each output state is likely due to the symmetry and balanced nature of the circuit. The Hadamard gates create superposition, resulting in an equal probability of the qubits being in the $\left|0\right\rangle$ or $\left|1\right\rangle$ state. The Controlled-Z gates introduce entanglement between qubits, but since the circuit is symmetric, the probabilities of the measured states remain balanced.

	\begin{table}
		\centering
		\begin{tabular}{ |p{2.5cm}||p{2.5cm}|p{1.5cm}|p{3cm}|p{2cm}|p{2cm}|p{2.5cm}| }
			\hline
			\multicolumn{7}{|c|}{Comparison between theoretical and experimental results} \\
			\hline
			Sender / receiver & Input Parameters & Phase shift Values & Input Probabilities & Qiskit-Qasm Simulator & Qiskit-Aer Simulator & Qiskit-Belem Simulator \\
			\hline
			Alice $\rightarrow$ Bob & $p=0$, $m=1$ & $\pi$ & [0.25, 0.25, 0.25, 0.25] & [0.249, 0.252] & [0.240, 0.251] & [0.257, 0.235] \\
			\hline
			Bob $\rightarrow$ Alice & $p=1$, $m=0$ & 0 & [0.25, 0.25, 0.25, 0.25] & [0.244, 0.253] & [0.245, 0.247] & [0.234, 0.259] \\
			\hline
		\end{tabular}
		\caption{Comparing Theoretical and Experimental Results: Probabilities of BQT of coherent state with varied parameters $p$ and $m$ in Phase Shift Gate using IBM-Qasm Simulator, IBM-Aer Simulator with 8192 shots for enhanced accuracy and quantum machine simulator (Belem Simulator), see Fig.(\ref{Fig6}).}\label{Tab2}
	\end{table}

We determine the probability of both even and odd states to facilitate a comparison between the experimental and theoretical results. The probability input is expressed as $P\left(x_{n}\right)=\vert\langle x_{n}\vert\psi\rangle\vert^{2}$, where $\vert\psi\rangle$ represents the initial coherent state. As shown in Table (\ref{Tab2}), we observe that the probability of measuring each even and odd state is approximately $0.25\%$. This indicates an equal chance of obtaining any of the possibilities "$0000$," "$0001$," "$1000$," and "$1001$" when implementing the BQT protocol. These probabilities serve as vital indicators of the effectiveness and efficiency of the BQT protocol and provide valuable insights for its optimization. To determine the angle of the phase shift, we multiply $m$ by $\pi$. We are able to accurately control the phase shift by changing the overlap parameter $p$ and the value of $m$ during the BQT in both directions and at different times. Evaluating the protocol's fidelity at various $p$ and $m$ values allows us to determine the specific combinations that result in the highest fidelity. We determine probabilities corresponding to these phase shift values using IBM simulators such as Qasm Simulator, Aer Simulator and Belem Simulator, allowing for a comparison between theoretically predicted and experimentally observed probabilities. This comparison analysis is critical in evaluating protocol performance and showing the agreement between theoretical predictions and implementation. The simulation results reveal intriguing insights into the BQT of coherent states. We show that the probabilities obtained using the Belem, Qasm, and Aer Simulators fit high fidelity values very well. Furthermore, these probabilities are very influenced by specific $p$ and $m$ values, where the fidelity varies significantly. The strong agreement between the probabilities and high fidelity values indicates the effectiveness of the teleportation operation. The output states are found to be very close to the initial quantum states, with a very few errors. The simulators generate results that closely approximate the ideal scenario of perfect teleportation, with maximum fidelity. Finally, our study shows the importance of probabilities in the context of coherent state teleportation and shows the implementation of the BQT protocol using entangled quantum channels. The combination of probabilities and high fidelity values on the Belem, Qasm, and Aer Simulators shows the protocol's efficacy in quantum teleportation. Our work yields intriguing results concerning the BQT of a coherent state, discovering that achieving perfect teleportation fidelity depends on the phase disparity between Alice's transmission to Bob and vice versa.

\section{Concluding remarks}
    To summarize, we proposed a bidirectional quantum teleportation scheme to simultaneously teleport even and odd coherent states in two directions through the multipartite Glauber coherent state. First, we examined the amount of entanglement in the shared resource state that can be maximally entangled, non-maximally entangled or separated depending on specific values of parameters such as overlap $p$, $m$ (symmetric or antisymmetric state) and the number of partitions $n$. We found that concurrence entanglement increases with higher values of overlap $p$ and reaches its maximum in the limiting case. This suggests that controlling the number of partitions and the overlapping of multi-partite coherent states can achieve pre-shared entanglement. Then, we evaluated the efficiency of the protocol via fidelity. We found that a minimal number of probes (e.g.$n=3$) is required for maximum fidelity. For the symmetrical state (i.e.$m=0$), fidelities in both directions reach their maximum when the shared resource is a entangled state ($p=1$). Conversely, for the anti-symmetrical state (i.e.$m=1$), fidelities increase and curve towards a maximum and in this situation, the state of the shared resource linking Alice to Bob corresponds to a Werner state ($p=1$). We have also observed that the fidelities reach their maximum when the shared resource is a maximally entangled state (GHZ-type with $p=0$ and $m=0$ or $p=0$ and $m=1$). Additionally, we compared the QFI with HSS associated with the trigger phase parameters and found that HSS plays a similar role to QFI in enhancing the efficiency of our BQT protocol in both directions. Then, we established a relationship between QFI and fidelity in both directions. In the case of the direction from Alice to Bob, we found that the variance of the estimated parameter is proportional to the fidelity. This implies that higher fidelity leads to a more accurate estimation of Alice's trigger phase parameter, denoted as $\vartheta_{e}$. It's important to note that this relationship is well-established in the literature within the context of standard quantum teleportation. Conversely, an inverse behavior was observed in the other direction, from Bob to Alice. In this scenario, lower fidelity actually results in a more accurate estimation of Bob's trigger phase parameter, $\vartheta_{o}$. These intriguing behaviors have been confirmed by Hilbert-Schmidt speed and offer valuable insights that go beyond the established knowledge of standard quantum teleportation. To determine the performance of the protocol and validate its predictions, probabilities obtained from IBM simulators are compared with the theoretically predicted probabilities. This enables a comparison between the expected and experimentally observed probabilities, providing insights into the agreement between theory and practice.\par
     
    Ultimately, maximizing and minimizing the fidelities of teleported states in both orientations depends on the Alice's and Bob's trigger states as well as the type of teleported information. Furthermore, the proposed BQT scheme is based on the multipartite coherent state as a resource generally influenced by environmental effects. So it is important to control the impact of the environment on the various entries parameters during the teleportation process, and to see how to realize this scheme in noisy channels as well as testing other quantum resource kinds beyond entanglement. We look forward to reporting on these issues in the next step.\par
 
{\bf Disclosures:} The authors declare no conflicts of interest or personal relationships that could have appeared to influence this work, and no Data associated in the manuscript.

\end{document}